\documentclass[11pt]{article}

\usepackage{color}
\usepackage[dvipsnames]{xcolor}
\usepackage[figuresright]{rotating}
\usepackage{natbib,graphicx,setspace,lscape,longtable,epstopdf,xr}
\usepackage{natbib,epsfig,graphicx, booktabs, multirow}
\usepackage{amsmath,amsthm,amssymb}
\usepackage{enumerate}
\usepackage{lmodern}
\usepackage{algorithm}
\usepackage{algpseudocode}
\usepackage[colorlinks,citecolor=blue,urlcolor=blue]{hyperref}
\usepackage{titlesec}
\usepackage{authblk} 
\usepackage{xr}
\usepackage{indentfirst}
\newtheorem{theorem}{Theorem}
\newtheorem{lemma}{Lemma}
\newtheorem{proposition}{Proposition}

\titleformat*{\section}{\centering\bf\large}
\titleformat*{\subsection}{\centering\it\large}

\def\one{{\bf 1}}
\def\bA{{\mathcal A}}
\def\ve{\varepsilon}

\def\tr{\mbox{tr}}

\def\beq{\begin{equation}}
\def\eeq{\end{equation}}
\def\beqr{\begin{eqnarray}}
\def\eeqr{\end{eqnarray}}
\def\beqrs{\begin{eqnarray*}}
\def\eeqrs{\end{eqnarray*}}
\def\bet{\begin{theorem}}
\def\eet{\end{theorem}}
\def\bel{\begin{lemma}}
\def\eel{\end{lemma}}
\def\bep{\begin{proposition}}
\def\eep{\end{proposition}}
\def\bg{\begin{figure}[tbph]\begin{center}}
\def\eg{\end{center}\end{figure}}

\def\bc{\begin{center}}
\def\ec{\end{center}}

\def\OS{\textup{os}}
\newtheorem{remark}{Remark}

\def\wt{\widetilde}

\def\wh{\widehat}

\def\ol{\overline }

\def\bB{\mathbf{B}}

\def\mN{\mathcal{N}}

\def\mD{\mathcal D}

\def\mR{\mathbb{R}}
\def\mI{\mathcal I}

\def\mS{\mathbb S}

\def\mS{\mathcal S}
\def\mE{\mathcal E}

\def\bM{\mathbb{M}}

\def\mX{\mathbb{X}}

\def\mY{\mathbb{Y}}

\def\var{\mbox{var}}

\def\cov{\mbox{cov}}

\def\diag{\mbox{diag}}

\newcommand{\bbeta}{\boldsymbol{\beta}}

\def\bX{\mathbf{X}}

\def\by{\mathbf{y}}

\def\bI{\mbox{\boldmath $I$}}

\def\zero{\mathbf{0}}
\def\defeq{\stackrel{\mathrm{def}}{=}}  

\newcommand{\bfm}[1]{\ensuremath{\mathbf{#1}}}

   \def\bA{\bfm A}  
   \def\bB{\bfm B}  
\def\bc{\bfm c}     
   \def\bD{\bfm D}  
\def\be{\bfm e}     
  \def\bF{\bfm F}  
\def\bg{\bfm g}     
     
   \def\bI{\bfm I}  
   \def\bJ{\bfm J}

   \def\bM{\bfm M}

   \def\bR{\bfm R}  
   \def\bS{\bfm S}  
   \def\bT{\bfm T}  
   \def\bU{\bfm U}  
\def\bv{\bfm v}   \def\bV{\bfm V}  
   \def\bW{\bfm W}  
   \def\bX{\bfm X}  
\def\by{\bfm y}     
     
\def\e{\bfm e}

\def\wh{\widehat}
\def\tR{\text{R}}

\def\boxit#1{\vbox{\hrule\hbox{\vrule\kern6pt\vbox{\kern6pt#1\kern6pt}\kern6pt\vrule}\hrule}}

\newcommand{\mylabel}[2]{#2\def\@currentlabel{#2}\label{#1}}

\newcommand{\bfsym}[1]{\ensuremath{\boldsymbol{#1}}}

 \def\bbeta{\bfsym \beta}

 \def\btheta{\bfsym {\theta}}           
           
              \def\bSigma{\bfsym \Sigma}

 \def\bpi{\bfsym {\pi}}
           \def\bXi{\bfsym {\Xi}}

\def\os {\textup{os}}

{\tiny }\textheight= 9in \textwidth = 6in

\numberwithin{equation}{section}

\begin{document}
	

\newcommand{\blind}{0}

\if0\blind
{\title{\bf \Large Distributed Estimation and Inference for  Spatial Autoregression Model with Large Scale Networks}
	\author[1]{Yimeng Ren $^*$}
	\author[1]{Zhe Li \thanks{Yimeng Ren and Zhe Li are joint first authors.}}
	\author[1,2]{Xuening Zhu \thanks{Xuening Zhu is the corresponding author. 
Email: \url{xueningzhu@fudan.edu.cn}.
}}
	\author[3]{Yuan Gao}
	\author[3]{Hansheng Wang}
    \affil[1]{\it \small School of Data Science, Fudan University}
	\affil[2]{\it \small MOE Laboratory for National Development and Intelligent Governance, Fudan University}
	\affil[3]{\it \small Guanghua School of Management, Peking University}
    \date{\vspace{-10ex}}
	\maketitle
} \fi

\if1\blind
{
	\title{\bf \Large Distributed Estimation and Inference for  Spatial Autoregression Model with Large Scale Networks}
	\bigskip
	\bigskip
	\bigskip
	\date{\vspace{-5ex}}
	\maketitle
	\medskip
} \fi

%
%
%
%


\begin{singlespace}
\begin{abstract}

The rapid growth of online network platforms generates large-scale network data and it poses great challenges for statistical analysis using the spatial autoregression (SAR) model. In this work, we develop a novel distributed estimation and statistical inference framework for the SAR model on a distributed system.
We first propose a distributed network least squares approximation (DNLSA) method.
	This enables us to obtain a one-step estimator by taking a weighted average of local estimators on each worker.
	Afterwards, a refined two-step estimation is designed to further reduce the estimation bias.
	For statistical inference, we utilize a random projection method to reduce the expensive communication cost.
	Theoretically, we show the consistency and asymptotic normality of both the one-step and two-step estimators.
In addition, we provide theoretical guarantee of the distributed statistical inference procedure.
	The theoretical findings and computational advantages are validated by several numerical simulations implemented on the Spark system. 
	Lastly, an experiment on the Yelp dataset further illustrates the usefulness of the proposed methodology.
	~

\end{abstract}
\end{singlespace}

\newpage

\section{INTRODUCTION}\label{sec:intro}

Consider a large-scale network with $N$ nodes, which are indexed as
$i = 1,\cdots, N$.
To characterize the network relationship among the network nodes, we employ an adjacency matrix $\bA = (a_{ij})\in\mR^{N\times N}$,
where $a_{ij} = 1$ implies that the $i$th node follows the $j$th node; otherwise, $a_{ij} = 0$.
Correspondingly, we collect an $N$-dimensional continuous response vector $\by = (Y_1,\cdots, Y_N)^\top\in\mR^{N}$ as well as the covariate matrix $\bX = (\bX_1,\cdots, \bX_N)^\top \in\mR^{N\times p}$.
To model the regression relationship among the nodes,
the spatial autoregression (SAR) model is widely used, and it is expressed as follows,
\beq
\by = \rho \bW \by + \bX\bbeta + \mE,\label{sar}
\eeq
where $\bW = (w_{ij})\in\mR^{N\times N}$ is the row-normalized adjacency matrix of $\bA$ with $w_{ij}
= n_i^{-1}a_{ij}$ and $n_i = \sum_j a_{ij}$.
In addition, $\mE = (\ve_1,\cdots, \ve_N)^\top\in\mR^N$ is the corresponding noise vector,
$\rho \in\mR$ and $\bbeta\in\mR^p$ represent unknown parameters to be estimated.

The SAR model as well as its extensions is widely applied to model data with observed network structures across a broad range of fields, which include
spatial data modeling \citep{lee2009spatial,shi2017spatial},
social behavior \citep{sojourner2013identification,liu2017peer,zhu2020multivariate},
financial risk management \citep{hardle2016tenet,zou2017covariance},
and many others.
Despite the usefulness of the SAR model, three main issues exist when applying it in practice.
First, when facing large-scale networks, while the estimation is feasible, it would take a high-end machine many days to obtain the results.
Second, the inference for the SAR model is difficult and even infeasible for large-scale networks, typically due to memory constraints and limited storage space.
Third, there are currently no available distributed algorithms that are well-established for the SAR model.
The above three issues have become increasingly important, especially in the era of big data.

To estimate the SAR model \eqref{sar}, a classical approach is to use the quasi-maximum likelihood (QMLE) method \citep{lee2004asymptotic}.
Although this approach is statistically efficient, the computational cost is extremely high because the inverse of a high-dimensional matrix $(\bI_N-\rho \bW)$ is involved in the estimation procedure \citep{huang2019least, zhu2020multivariate}.
To reduce the computational burden, the IV-based methods, such as the two stage least squares (2SLS) estimation
and three stage least squares (3SLS) estimation methods, have also been developed and are widely used \citep{kelejian2004estimation,baltagi2015ec3sls,cohen2018multivariate}.
However, the implementation of these methods relies on exogenous variables.
If ideal exogenous variables are not available, such estimation methods are less flexible.
Recently, \cite{huang2019least} and \cite{zhu2020multivariate} propose estimating the SAR model by constructing a novel least squares (LS) type objective function.
This approach takes advantage of the network's sparsity structure to reduce the computational complexity, which is desirable for large-scale network data.

\begin{figure}[htpb!]
	\centering
	\includegraphics[width=0.4\linewidth]{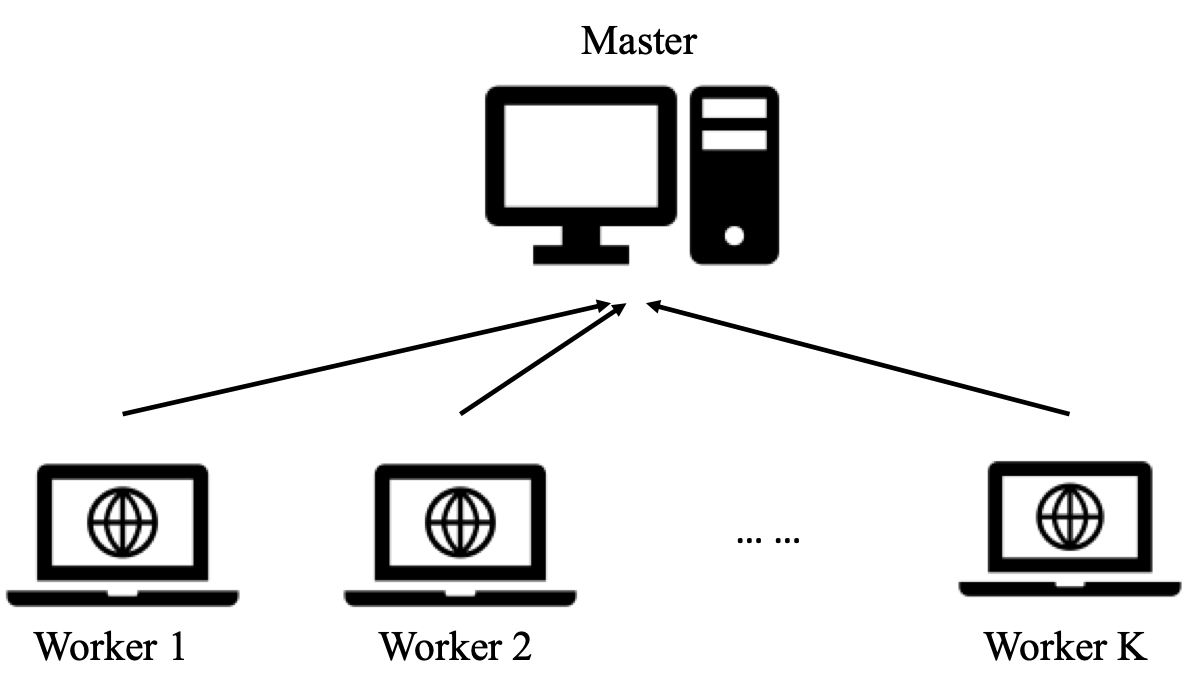}
	\caption{\small Illustration of distributed system. A distributed system consists of multiple workers and a single master computer.}
	\label{fig:dsar}
\end{figure}

Although the above mentioned approach is useful for handling large-scale network data on a single computer, it is not scalable for a distributed system.
Besides, conducting the statistical inference involves more complicated calculations, which makes it even infeasible with large-scale networks, since it is usually restricted by the memory constraint and the requirement for large storage space.
This makes the statistical inference in a distributed system to be a more preferable and feasible choice for large networks.
To better distribute computing tasks for large-scale dataset, a typical ``workers-and-master" type distributed system has been considered and adopted by popularly used distributed environments such as Hadoop \citep{dean2004mapreduce} and Spark \citep{zaharia2010spark}.
In this system, the master and all of the workers are modern computers with reasonable computing power and storage capacity.
According to Figure \ref{fig:dsar}, applying the distributed system for a single round of communication generally requires three steps. First, the whole mission is divided by the master and allocated to each worker.
Second, all of the workers execute the sub-task with the local dataset and transmit the results to the master.
Finally, the results are integrated by the master to generate the final result.
During the whole process, there is no communication among workers; hence, the total time cost is composed of only the worker computing time, the master reducing time and the worker-master communication time.
We remark that the communication cost can be important when designing a distributed algorithm \citep{jordan2018communication,XiangyuChang2017DivideAC,fan2019distributed, chen2020distributed,fan2021communication}.
The communication cost refers to the wall-clock time cost needed for data communication between different computer nodes \citep{zhu2021least}, which is mainly determined by two factors.
The first factor is the number of communication rounds for a distributed system.
In this regard, fewer rounds of communication are preferred to save costs \citep{jordan2018communication,fan2019distributed}.
The second factor is the amount of transmitted data between the workers and the master during each round of communication.
In this regard, smaller sizes of transmitted data are preferred to save costs.

To accomplish the distributed estimation of the SAR model, we face two main challenges.
The first challenge is how to design the distributed strategy of the network data in a distributed system.
In the existing literature, the data are usually distributed by splitting samples (i.e., rows) \citep{jordan2018communication, fan2019distributed} or features (i.e., columns) \citep{smith2018cocoa, li2020distributed}.
However, for network data, these strategies would break the network dependency inside the data stored on different nodes.
Besides, the simple ``divide-and-conquer" type algorithm \citep{zhang2013communication,liu2014distributed,lee2015communication,battey2015distributed,fan2019distributed} cannot be directly applied. 
Namely, if we simply divide the samples into $K$ sub-samples, and then conduct the SAR model estimation  based on local data and the sub-network relationships,
the resulting estimator would be inconsistent \citep{chen2013impact, zhou2017estimating}.
The second challenge is how to combine the local estimators to produce the final estimator.
 If we take simple average of the local estimators, the estimation efficiency will be barely satisfactory \citep{zhu2021least}.
Consequently, how to conduct local computation and design an ideal combination strategy to yield the final estimator becomes an important problem.

To address the above two issues, we propose a distributed least squares estimation for the SAR model in a distributed system.
The idea is motivated by both the least squares estimation (LSE) method \citep{huang2019least,zhu2020multivariate}
and a recently proposed distributed least squares approximation (DLSA) method \citep{zhu2021least}.
As suggested by the LSE method, the network effect can be consistently estimated for a sub-network as long as the nodes and their connected friends up to a second-order connection are contained in the sub-network.
Specifically, the calculation of the LSE only involves the first-order and a certain kind of second-order friends of the interested nodes.
The sub-network details are stated in Section \ref{subsec:LSE_for_sar}.
Therefore, the estimation can be computationally efficient especially when the network is sparse.
This motivates us to assign a local network on each worker to obtain a consistent local estimator in a distributed system.
Subsequently, a major problem is how to aggregate the local estimators on the master computer.
A straightforward solution is to take simple average of the local estimators to yield the final estimator,
which is typically referred to as ``one-shot" (OS) estimation in literature \citep{zhang2013communication, battey2015distributed, XiangyuChang2017DivideAC}.
Although it can yield a consistent estimator,
however, it is suboptimal compared to the global estimator which uses the whole network information.
To solve this problem, we borrow the wisdom of the DLSA method \citep{zhu2021least} to approximate the objective function with local quadratic functions.
This enables us to obtain an analytical formula to aggregate the local estimators on the master computer.
Despite the similarity with the DLSA method, our analysis is based on the network dependent data setting, while they focus their study on the independent and identically distributed data.
We refer to the proposed method as distributed network least squares approximation (DNLSA) method.
Further theoretical investigation shows that the resulting estimator can achieve the global estimation efficiency as using the whole network data.
In addition, the communication cost is carefully controlled.
Moreover, to reduce the estimation bias, we refine the one-step estimator with an additional estimation step, which leads to a two-step estimator.
This can allow even smaller local sample sizes and retain desirable performances.


Further, despite the useful strategy of the distributed estimation for the SAR model, we still confront another critical challenge when conducting statistical inference.
The main difficult is that the local network data on each computer are not independent,
hence the DNLSA method cannot allow for direct distributed statistical inference.
Detailed investigation shows that it requires
each worker to communicate an $N\times N$ dimensional matrix  to the master, in order to exactly estimate the asymptotic covariance matrix.
The transferred data size for this method is $O(N^2)$,
which is extremely expensive for large-scale networks.
To reduce the communication cost, we propose a random projection method for distributed statistical inference.
Specifically, we use random matrices to project the matrix of $N\times N$ dimension to a much lower dimension, i.e., $d\times d$.
Then we transmit the low dimensional matrix from workers to the master.
{This substantially decreases the communication costs, as the transferred data size is effectively reduced from  $O(N^2)$ to $O(d^2)$.}
Our theoretical investigation suggests that setting $d\ge c\log N$ ($c>0$) is sufficient to obtain a consistent
estimator for the asymptotic covariance matrix.
This makes the distributed statistical inference feasible with low communication cost.


The rest of the article is organized as follows.
Section \ref{sec:model} introduces the SAR model and the DNLSA algorithm, as well as the theoretical analysis.
In Section \ref{sec:inference},  we develop a random projection method to facilitate the distributed inference.
Multiple simulation studies are provided in Section \ref{sec:num}, and a real data application is illustrated by applying the DNLSA method on the Spark system in Section \ref{sec:realdata}.
In Section \ref{sec:conclusion}, we briefly summarize the article and make a concluding remark. All the technical details, theoretical proofs {and additional numerical results} are elaborated in the Appendices.


\section{DISTRIBUTED ESTIMATION FOR THE SAR MODEL}\label{sec:model}

\subsection{Least Squares Estimation for the SAR Model}\label{subsec:LSE_for_sar}

We first provide a brief introduction to the SAR model, which is originally proposed to analyze spatial data \citep{ord1975estimation, lee2003best, lee2004asymptotic}.
The vector form of the SAR model is expressed in \eqref{sar} as follows,
\begin{align*}
Y_i = \rho \sum_j w_{ij} Y_{j} + \bX_i^\top\bbeta +\ve_i, ~~~ i = 1, \cdots, N.
\end{align*}
Spatial data analysis assumes that the observation in the $i$th location can be modeled as a weighted average of its spatial neighbors, its own covariates and random noise.
Consequently, it characterizes the spatial dependence structure among the spatial regions.
Recently, the SAR model has gained popularity for modeling network data since it shares many similarities with spatial data.
For instance, in social network analysis, the observations can be activity measurements collected from network users, and the adjacency matrix $\bA$
is defined by the following-followee relationship \citep{zhu2017network, huang2019least, wu2022inward}.
In this regard, $\rho$ is typically referred to as the network effect.
Because the term $\sum_j w_{ij} Y_{j}$ is correlated with $\ve_i$, we have an endogeneity issue for estimation, and various estimators are proposed in the literature \citep{kelejian1998generalized,lee2003best, baltagi2011maximum, baltagi2015ec3sls}.
Since we are considering a large-scale network analysis problem, we employ the LSE method, which is a framework recently proposed by \cite{huang2019least}, to reduce the computational burden.

\begin{figure}[htpb!]
	\begin{center}
		\includegraphics[width=0.6\textwidth]{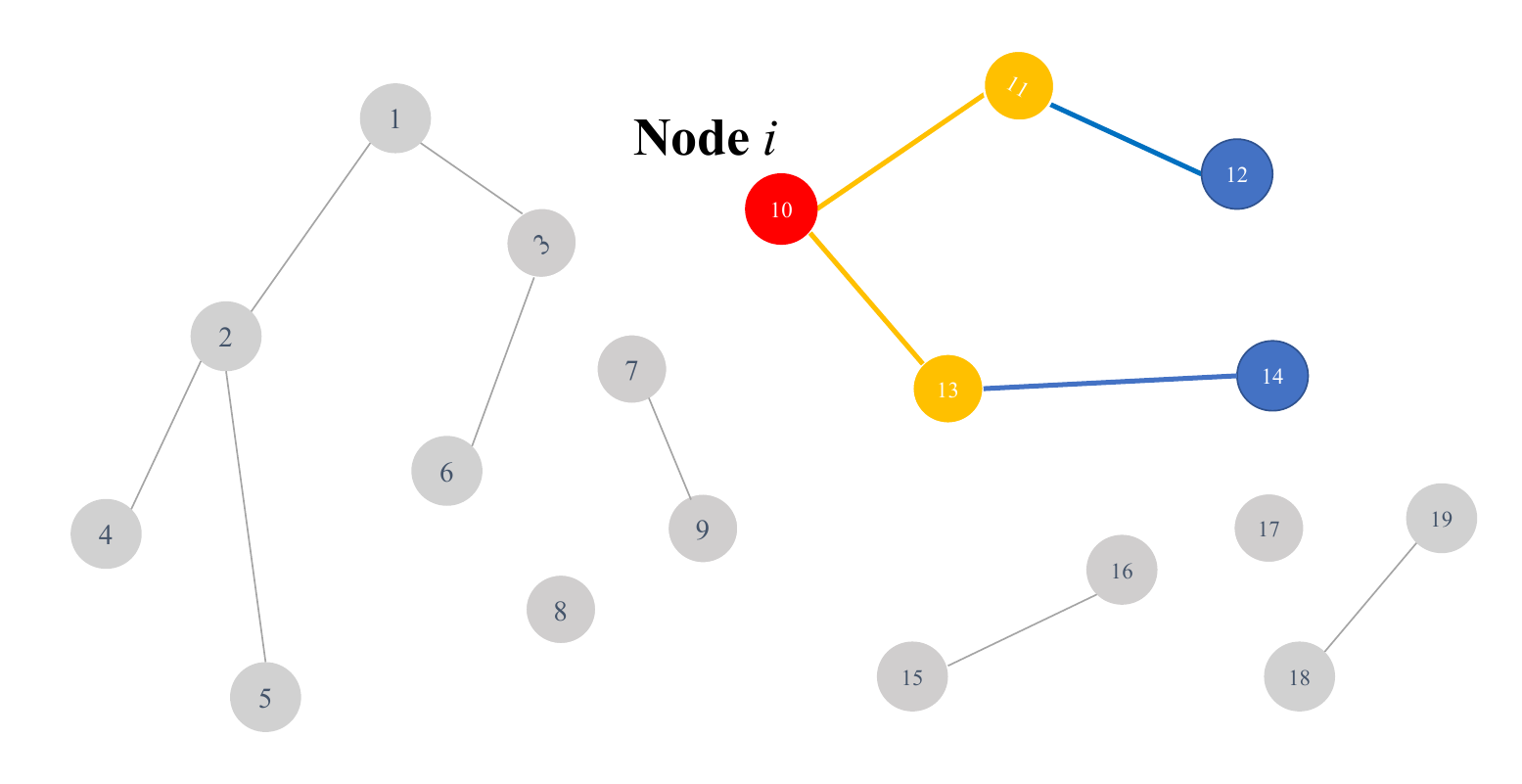}
		\caption{\small First and second-order friends of node $i$.
For a node $i$ in the network {(marked in red)}, its first-order friends are marked in yellow and its second-order friends are marked in blue.}
		\label{fig:node_j}
	\end{center}
\end{figure}

Since our distributed algorithm for the SAR model is motivated by the LSE method proposed by
\cite{huang2019least} and \cite{zhu2020multivariate},
we first introduce the basic idea of the LSE method.
Let $\mY_{-i} = (Y_j, j\ne i)^\top$ collect the responses of all nodes except for the $i$th node.
Suppose $\mE$ follows multivariate normal distribution $N(\zero, \sigma_\ve^2I_N)$ at this moment.
Denote $\btheta = (\rho, \bbeta^\top)^\top\in\mR^{p+1}$ as the parameter of interest.
It is easy to verify that
$
\wt Y_i(\btheta) = E\big\{Y_i|\mY_{-i}\big\} = \mu_i+ \sum_{j\ne i} \alpha_{ij}(Y_j - \mu_j),
$
where
\beq
\alpha_{ij} = \frac{\rho(w_{ij}+w_{ji}) - \rho^2\sum_k w_{ki}w_{kj}}{1+\rho^2\sum_k w_{ki}^2}\label{weight}
\eeq
and $\mu_i = E(Y_i)$.
The detailed derivation can be found in Section 2 of the supplementary material of \cite{zhu2020multivariate}.
As a consequence, the conditional expectation $E\{Y_i|\mY_{-i}\}$ can be written as a linear combination of the other responses.
Inspecting (\ref{weight}),
one can find that for the $i$th node, the weights are related to its
first- and second-order network relationships.
Namely, the first-order friends are collected by $\{j: w_{ij}\ne 0\mbox{ or }w_{ji}\ne 0\}$,
and the second-order friends are collected by $\{j: \sum_k w_{ki}w_{kj}\ne 0\}$.
In particular, Figure \ref{fig:node_j} depicts the first- and second- order friends
of a node $i$ in the network.
If the network structure $\bW$ is sufficiently sparse,
then the number of nodes involved in computation is small.
Hence, the total computational burden can be reduced.

Based on the conditional expectation, we con construct an LS type objective function as follows,
\beq
Q(\btheta) = \frac{1}{N} \sum_i \big|Y_i - \wt Y_i(\btheta)\big|^2 = \frac{1}{N}\big\|\bD\bS^\top \{\bS\by - \bX\bbeta\}\big\|^2 \defeq \frac{1}{N}\bF(\btheta)^\top \bF(\btheta),\label{obj_func}
\eeq
where $\bF(\btheta) = \bD\bS^\top \{\bS\by - \bX\bbeta\}$ and,
\begin{align}
	\bD =  \{\bI+\rho^2 \diag(\bW^\top \bW)\}^{-1},
	~~~\mbox{and}~~ \bS = \bI - \rho \bW\label{DS}.
\end{align}
The derivation from \eqref{weight} to \eqref{obj_func} can be found in \ref{subsec:object_func_derive}.
Note that the above objective function does not involve
the inverse of a high dimensional matrix $\bI-\rho \bW$ as in the QMLE method \citep{lee2004asymptotic}.
Consequently, the computational complexity will be largely reduced.
We further remark that although the LS method is motivated by the assumption that $\mE$ follows a normal distribution, the method is still feasible
for the non-normal case.
We refer to \cite{huang2019least} and \cite{zhu2020multivariate} for comprehensive discussions, and in the following section, we introduce a distributed algorithm for the SAR model based on the least squares estimation method.

Throughout the rest of this paper, the cardinality of a set $\mS$ is denoted
by $|\mS|$. We use $I(\cdot)$ to denote the indicator function.
For a vector $\bv = (v_1,\cdots, v_p)^\top\in\mR^p$, define
$\|\bv\|_q = (\sum_{j = 1}^p v_j^q)^{1/q}$ for $q>0$.
For convenience we omit the subscript $q$ when $q = 2$.
For an arbitrary matrix $\bM = (m_{ij})\in\mR^{p_1\times p_2}$,
denote $\|\bM\|_F = \tr(\bM^\top \bM)^{1/2}$ as the Frobenius norm.
Here, we use $\tr(\cdot)$ as the trace of a square matrix.
For a square symmetric matrix, we use $\lambda_{\min}(\cdot)$ and $\lambda_{\max}(\cdot)$ to denote their smallest and largest eigenvalues, respectively.
Similarly, $\sigma_{\min}(\cdot)$ and $\sigma_{\max}(\cdot)$ represent the smallest and largest singular values.
For a matrix $\bM = (m_{ij}) \in \mR^{p_1\times p_2}$, denote $\|\bM\|$
as its largest singular value.
Let $\bM^{(\mS,\cdot)} = (m_{ij}: i\in \mS, 1\le j\le p_2)\in \mR^{|\mS|\times p_2}$ and
$\bM^{(\cdot,\mS)} = (m_{ij}: 1\le i\le p_1, j\in \mS)\in \mR^{p_1\times |\mS|}$ be sub-matrices of $\bM$.
For two arbitrary sequences $\{a_N\}$ and $\{b_N\}$,
$a_N\gtrsim b_N$ implies that there exists a positive constant $c$ and $N_0 > 0$, such that $a_N \ge c b_N$ for any $N > N_0$.
We also define $a_N \gg b_N$ as $a_N/b_N \to \infty$ as $N \to \infty$.
Lastly, we use $\e_i$ to denote the $i$th unit vector {of length $N$}, with the $i$th element being 1 and the others being 0.

\subsection{Distributed Least Squares Estimation with Local Network}\label{subsec:estimation}

\begin{figure}[htpb!]
	\begin{center}
		\includegraphics[width=0.8\textwidth]{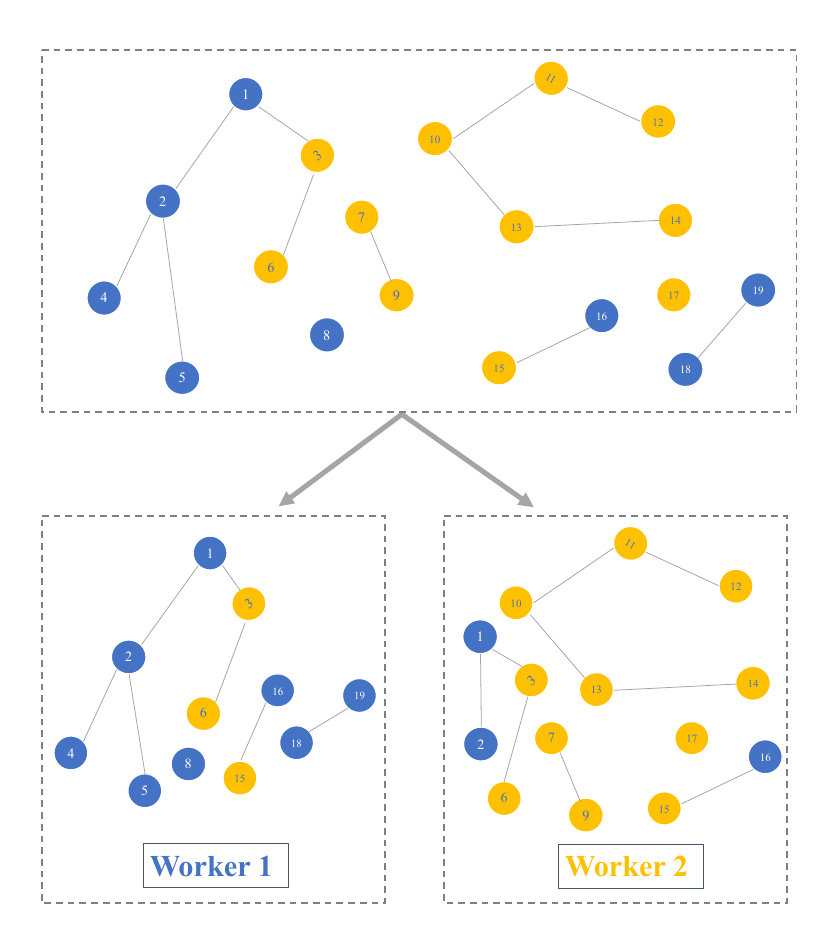}
		\caption{\small Local storage scheme for a network with $K=2$. The blue nodes with their first- and second-order friends' information are stored on worker 1, and the yellow nodes with up to second-order friends' information are stored on worker 2.}
		\label{fig:local_store}
	\end{center}
\end{figure}

It is noteworthy that estimation by optimizing the objective function (\ref{obj_func}) only involves  the first- and second-order network relationships of each node $i$, which motivates us to propose an LS-based distributed algorithm for the SAR model estimation.
We refer to this method as the DNLSA algorithm.
Suppose the $N$ nodes are distributed on $K$ workers, and $\mS = \{1,\cdots, N\}$ is defined as the index set of all nodes.
Correspondingly, let $\mS_k$ be the set of nodes on the $k$th worker and $N_k = |\mS_k|$ be the number of nodes on this worker.
Similarly, we define the objective function on each worker as follows,
\beq
Q_k(\btheta) = \frac{1}{N_k} \sum_{i \in \mS_k} |Y_i - \wt{Y}_i(\btheta)|^2\label{Q_func}
\eeq
Then, we have
\beq
Q(\btheta) = \frac{1}{N} \sum_k N_k Q_k(\btheta) = \sum_k \alpha_k Q_k(\btheta)
\eeq
where $\alpha_k = N_k/N$.
Recall from (\ref{obj_func}) that we can write $Q(\btheta) = N^{-1}\bF(\btheta)^\top \bF(\btheta) \defeq N^{-1} \sum_i F_i(\btheta)^2$, where
\begin{align*}
	F_i(\btheta) &=\e_i^\top \bF(\btheta) = \e_i^\top \bD(\bI-\rho \bW)^\top \big\{(\bI-\rho \bW)\by - \bX\bbeta\big\}\\
	& = \e_i^\top \bD(\bI-\rho \bW)^\top (\bI-\rho \bW)\by - \e_i^\top \bD(\bI-\rho \bW)^\top \bX\bbeta.
\end{align*}
Define $\wh \btheta_k = \arg\min_{\btheta} Q_k(\btheta)$ as the local estimator on worker $k$.
To obtain $\wh \btheta_k$, we write $Q_k(\btheta) $ as
$Q_k(\btheta) = N_k^{-1}\sum_{i\in\mS_k} F_i(\btheta)^2$.
Then, it is crucial to calculate $F_i(\btheta)$ on the worker.
Specifically,
to compute $F_i(\btheta)$
for the $i$th node on the $k$th worker, it requires calculating
$ \wt d_i = \bW_{\cdot i}^\top \bW_{\cdot i}$,
$ \bW_{i\cdot}\by = \sum_{j = 1}^Nw_{ij}Y_j$,
$\bW_{\cdot i}^\top \by = \sum_{j = 1}^N w_{ji} Y_j$,
$ \bW_{\cdot i}^\top\bX = \sum_{j = 1}^N w_{ji} \bX_{j}^\top$
and $\bW_{\cdot i}^\top \bW \by = \sum_{j = 1}^N w_{ji}^{(2)} Y_j$,
where
$w_{ji}^{(2)} = \sum_{k = 1}^N w_{ki} w_{kj}$.
Note that $w_{ij}$ ($w_{ji}$) and
$w_{ji}^{(2)}$ are the first-order and one of the second-order network relationships of the local node $i$.
To provide a better understanding,
we refer to the node sets $\mN_i^{out} = \{j: w_{ij}\ne 0\}$ and
$\mN_i^{in} = \{j: w_{ji}\ne 0\}$
as the local-out-network and local-in-network, respectively.
In addition,
we refer to the set $\mN_i^{(2)} = \{j: w_{ji}^{(2)}\ne 0\}$ as the local-second-order-network for $i$.
As a result, to compute $ F_i(\btheta)$, we need to store the following local network information of node $i$: (a)
the value $\wt d_i$;
(b) the averaged node responses from local networks $\mN_i^{out},
\mN_i^{in}, \mN_i^{(2)}$, i.e.,
$\sum_{j\in \mN^{out}} w_{ij}Y_j$, $\sum_{j \in \mN^{in}} w_{ji} Y_j$ and $\sum_{j \in \mN^{(2)}} w_{ji}^{(2)} Y_j$; and
(c) the averaged node covariates from $\mN_i^{in}$, i.e.,
$\sum_{j\in \mN_i^{in}} w_{ji} \bX_{j}^\top$.
As a consequence, instead of directly dividing the whole network structure, we actually need to store a local sub-network on each worker.
For illustration, Figure \ref{fig:local_store} shows how the sub-network information related to $\mS_k$
is stored on worker $k$ for $K= 2$ workers in total.
As shown in Figure \ref{fig:local_store}, some nodes may be duplicated stored in sub-networks on each worker.
This is related to how nodes are assigned on each worker.
We discuss the storage requirement and computational cost under a stochastic block network in \ref{subsubsec:simu_storage} under different nodes assignment strategies.
For a sparse network, the local network sizes should be small and thus,
the local computational cost can be controlled.

Next, to conduct the distributed estimation of the SAR model, a
straightforward method is to take a simple average of the local
estimators $\wh \btheta_k$, which is typically referred to as one-shot (OS) estimator in the literature
\citep{zhang2013communication, battey2015distributed}.
Specifically, denote the OS estimator as $\wh\btheta^{\OS} = K^{-1} \sum_k \wh\btheta_k$.
Despite its simple form, this estimator is not necessarily globally efficient \citep{zhu2021least, cai2021individual} due to the heterogeneous local information
across different workers.
Consequently, to achieve global efficiency, we decompose and approximate the global objective function around the local estimators
by using a local quadratic form as follows,
\begin{align}
	Q(\btheta) & = \sum_{k = 1}^K \alpha_k Q_k(\btheta)  = \sum_{k = 1}^K  \alpha_k  \Big\{Q_k(\btheta) - Q_k(\wh \btheta_k)\Big\} + C_1\nonumber\\
	& \approx \sum_{k = 1}^K  \alpha_k  (\btheta - \wh \btheta_k)^\top \ddot Q_k(\wh \btheta_k) (\btheta - \wh \btheta_k) + C_2,\label{eq:Q_theta}
\end{align}
where $\ddot{Q}_k(\btheta)$ is the second-order derivative of $Q_k(\btheta)$.
Here, $C_1$ is only related to $\wh \btheta_k$ and $C_2$ contains higher order expansion at $\btheta$, which is omitted here.
The ``$\approx$" in \eqref{eq:Q_theta} is used to keep only the main quadratic term.
This implementation motivates us to define the following weighted least squares type loss function,
\beq
Q^w(\btheta) = \sum_{k = 1}^K \alpha_k (\btheta - \wh \btheta_k)^\top \ddot Q_k(\wh \btheta_k) (\btheta - \wh \btheta_k).\label{sur_loss}
\eeq
By minimizing the above surrogate objective function, we can obtain the following weighted least squares estimator (WLSE),
\beq
\wh \btheta^{w} = \Big\{\sum_{k = 1}^K \alpha_k \ddot Q_k(\wh \btheta_k)\Big\}^{-1} \Big\{\sum_{k = 1}^K \alpha_k \ddot Q_k(\wh \btheta_k) \wh \btheta_k\Big\}.\label{wlse}
\eeq
As implied by \eqref{wlse}, one only needs one round of communication to obtain the WLSE.
First, each worker conducts a local computation and produces the local estimator $\wh \btheta_k$.
Second, we transmit $\wh \btheta_k$  and $\ddot Q_k(\wh \btheta_k)$ from the workers to the master to obtain the final WLSE by \eqref{wlse}.
Theoretically, it is interesting to investigate whether the statistical efficiency of the WLSE could match the global estimator $\wh \btheta = \arg\min_{\btheta} Q(\btheta)$, and we present the details in the next section.

\subsection{Theoretical Properties}\label{subsec:theory}

Denote the true parameter as $\btheta_0 = (\rho_0, \bbeta_0^\top)^\top$.
To facilitate the theoretical discussions, we first present the following technical conditions.



\begin{enumerate}[(C1)]
	\item\label{con:noise}(\textsc{Noise Term}) The random errors $ \varepsilon_1,\dots, \varepsilon_N $ are independent and identically distributed random noise with zero mean, and follow a sub-Gaussian distribution, such that $ E\{\exp(t\varepsilon_i)\} \le e^{a^2t^2/2}$ for some positive constant $a>0$ and $t>0$.
In addition, assume that $E(\ve_i^3) = 0$.

	\item\label{con:cov}(\textsc{Covariates})
		Let $\bM$ be an $N \times N$ dimensional matrix. Suppose that $\| \bM \|_F = O(k_N)$ where $k_N \to \infty$ as $N \to \infty$. Assume $k_N^{-1} |(\bX \bbeta_0)^\top \bM (\bX \bbeta_0)| \le c_x k_N^{-1} |\tr\{ \bM \}|$
		as $N \to \infty$, where $c_x$ is a finite positive constant.
	Further assume that $\|\bX_i \| \le c_x$, where $c_x$ is a positive constant.

	\item\label{con:network} (\textsc{Network Structure})
	\begin{enumerate}[(C3.1)]
		\item\label{con:connect} (\textsc{Connectivity})  Assume that the set of all nodes $\mS = \{1, \cdots, N\}$ is the state space of an irreducible and aperiodic Markov chain. The transition probability is expressed as $\bW$. Define $\bpi = (\pi_1, \cdots, \pi_N)^\top \in \mR^N$ as the stationary distribution vector of the Markov chain (i.e., $\bW^\top \bpi = \bpi$) with elements $\pi_i \ge 0$ and $\sum_i \pi_i = 1$.
			Suppose that $ \sum_{i=1}^N \pi_i^2 = O(N^{-1/2 - \delta}) $, where $0<\delta\le 1/2  $ is a positive constant.
%
		\item\label{con:uniform}(\textsc{Uniformity})  Denote $\bW^* = \bW+\bW^\top$, and assume that $ |\lambda_{\max}(\bW^*)| = O(\log N) $.
	\end{enumerate}
	
	\item\label{con:param}{(\textsc{Parameter Space})} Assume $ \btheta \in \Theta $, where $ \Theta $ is a compact and convex subset of $ \mR^{p+1} $. In addition, the true value $ \btheta_0 $ lies in the interior of $\Theta$.
	
	\item\label{con:sample_size} (\textsc{Local Sample Size})
	Let $n = N/K$ and suppose $c_1 \le \min_k N_k/n \le \max_k N_k/n \le c_2$ for some positive constants $c_1$ and $c_2$.

	\item\label{con:ident} (\textsc{Identification Condition})
Denote $\mI_k = (\be_i: i \in \mS_k)^\top \in \mR^{N_k \times N}$ and  $ {\mX}_k = (  \mI_k \bD_0 \bS_0^\top \bW \bS_0^{-1} \bX \bbeta_0,  \mI_k \bD_0 \bS_0^\top\bX) \in \mR^{N_k \times (1+p)}$, where
$\bD_0 $ and $\bS_0$ are the true values of $\bD$ and $\bS$ in \eqref{DS} by substituting $\btheta_0$.
	Assume that $N_k^{-1} \lambda_{\min} ( {\mX}_k^\top  {\mX}_k)>c_0$ for all $1\le k\le K$ as $N_k\rightarrow\infty$, where $c_0$ is a positive constant.
	
		\item\label{con:converge} (\textsc{Convergence})
		Define $\bSigma_{1kl} = \sqrt{N_k N_l}\cov\{\dot Q_k(\btheta_0), \dot Q_l(\btheta_0)\}$ and $\bSigma_1 = \sum_{k,l=1}^K \sqrt{\alpha_k \alpha_l} \bSigma_{1kl}$,
		where $\dot{Q}_k(\btheta_0)$ is the first order derivative of $Q(\btheta_0)$.
		The analytical forms of $\bSigma_{1kl}$ are provided in \ref{asymp_cov}.
		Assume $\lambda_{\min}(\bSigma_{1})\ge \tau_0$ and  $\max_{k,l}\sigma_{\max}(\bSigma_{1kl}) \le  \tau_1$, where $\tau_0$ and $\tau_1$ are two positive constants.
\end{enumerate}

We comment on the conditions in  the following. First,
Condition (C\ref{con:noise}) assumes that the noise term follows the sub-Gaussian distribution, which is a milder assumption than the normal distribution.
It is widely used in high dimensional modeling literature \citep{negahban2011estimation, negahban2012unified, jordan2018communication}.
Subsequently, Condition (C\ref{con:cov}) can be regarded as a law of large number type assumption about the covariates.
The same type of condition can be found in \cite{zhu2021network}.
Both (C\ref{con:noise}) and (C\ref{con:cov}) facilitate asymptotic analysis and the adoption of the central limit theorem.

Condition (C\ref{con:network}) imposes assumptions on the network structure, which include two separate parts. Condition \hyperref[con:connect]{(C3.1)} assumes a certain connectivity for the network structure.
This condition assures that any two nodes in the network can be connected with a finite number of steps.
For real-world networks, this condition can be easily satisfied \citep{newman2006modularity}.
Otherwise, the whole network can be decomposed into a number of fully separated sub-networks, and each sub-network should be modeled separately.
Condition \hyperref[con:uniform]{(C3.2)} allows $\lambda_{\max}(\bW^*)$ to diverge at a rate of $O(\log N)$.
This implies that the node's degrees can diverge as $N\rightarrow\infty$ but at a slower rate.
Compared to the bounded assumption on the column sums of $\bW$  \citep{lee2010estimation,tao2012spatial, yang2016bias},
our assumption is milder and more natural in the network data setting.
In addition, as implied by Condition (C\ref{con:network}), we actually do not necessarily need $a_{ij}\in \{0,1\}$ as long as the weight matrix $\bW$ satisfies (C\ref{con:network}).

Subsequently, Condition (C\ref{con:param}) assumes the parameter space to be compact \citep{jordan2018communication}, and
Condition (C\ref{con:sample_size}) assumes that the local sample sizes diverge at the same speed to facilitate the theoretical discussions.
Next, Condition (C\ref{con:ident}) is an identification assumption imposed on the matrix $\mX_k$.
This assumption is similar to the identifiability condition in
\cite{zhu2020multivariate} but uses the sub-network information on the $k$th worker (i.e., $\mS_k$)
under the distributed data setting.
Lastly, Condition (C\ref{con:converge}) ensures the convergence of the corresponding
matrices, and similar conditions have been imposed by \cite{jordan2018communication} and \cite{zhu2021least}.

Given the above conditions, we start with the asymptotic bias-variance analysis of the estimator $\wh \btheta^w$.
This provides us with important insights to further establish the asymptotic normality result.

\begin{proposition}[\textsc{Bias-Variance Analysis}] \label{prop.1}
	Assume conditions (C\ref{con:noise})--(C\ref{con:converge}) hold and $K = O(N^\zeta) \ (0<\zeta<1)$.
	Then we have $\sqrt{N}(\wh \btheta^{w} - \btheta_0) = {\wh  \bSigma_2^{-1}}\{{\bV}(\btheta_0) + {\bB}_1(\btheta_0)\}$, where $\wh  \bSigma_2 = \sum_{k = 1}^K \alpha_k \ddot Q_k(\wh \btheta_k)$,
$E\{\bV(\btheta_0)\} = \zero$, $\|\cov\{\bV(\btheta_0) \}\|_F = O(1)$, and $\|\bB_1(\btheta_0)\| = O_p\{ K (\log N)^{8}/\sqrt{N} \} $.
In addition, we have $\wh \bSigma_2 \to_p \bSigma_2$, where $\bSigma_2 = \sum_k \alpha_k \bSigma_{2k}$ and $\bSigma_{2k} =  E\{\ddot{Q}_k(\btheta_0)\}$ is given in \ref{asymp_cov}.
\end{proposition}

The proof of Proposition \ref{prop.1} is provided in \ref{append.prop1}.
Proposition \ref{prop.1} separates $\sqrt{N}(\wh \btheta^{w} - \btheta_0)$ into two parts, namely, the variance part and bias part.
Particularly, the variance part is not related to $K$  but the bias part is.
When the number of workers $K$ increases, the local sample size $N_k$ drops down, then the bias order becomes larger, while the variance term remains the same.
A similar conclusion is obtained by distributed estimation under the independent data setting \citep{zhu2021least}.
Compared to the result in the independent data setting, we note that the bias order under our setting is slightly higher.
That is because network dependence is involved in our asymptotic analysis.
To make the asymptotic bias ignorable (i.e., $ \|\bB_1(\btheta_0)\| = o_p(1)$), we need
$K\ll \sqrt N/(\log N)^8$, which is equivalent to
assuming that the local sample size is $n \gg N^{1/2} (\log N)^8$, which is a slightly higher requirement for the local sample size than that of the independent data setting.
Subsequently,
we establish the following asymptotic normality result.


	\begin{theorem}[\textsc{Asymptotic Normality for WLSE}] \label{thm.1}
		Assume Conditions {(C\ref{con:noise})--(C\ref{con:converge})}, then we have $\sqrt{N}(\wh \btheta^{w} - \btheta_0)\to_d N(\zero, {\bSigma_2^{-1}\bSigma_1\bSigma_2^{-1}})$ if $n/\{N^{1/2} (\log N)^{8}\}\to \infty$, where $\bSigma_1 = \sum_{k, l = 1}^K \sqrt{\alpha_k \alpha_l} \bSigma_{1kl}$.
	\end{theorem}
	
The proof of Theorem \ref{thm.1} is provided in \ref{append.thm1}.
The condition $n/\{N^{1/2} (\log N)^{8}\}\to \infty$ is used to guarantee that the asymptotic bias can be ignored.
This approach motivates us to consider further reducing the bias to refine the one-step estimator; thus, we can allow smaller local sample sizes.
To this end, we propose a refined two-step estimation method in the next section.

\subsection{A Refined Two-Step Estimation}\label{subsec:twostep}
	
	We note that Theorem \ref{thm.1} requires that $n/\{N^{1/2} (\log N)^{8}\}\to \infty$. This is an assumption that may be violated if the local sample size is insufficient.
For instance, if we are available to a large number of workers (i.e., large $K$),
we will have a smaller local sample size $n$ on each worker, which implies that
the condition $n/\{N^{1/2} (\log N)^{8}\}\to \infty$ may be violated.
To relax the restriction on the local sample size, we next propose a two-step WLSE (TWLSE) to refine our previous one-step estimator $\wh{\btheta}^w$.
The basic idea is to use one additional iteration to conduct the estimation.
This will consume one more round of communication but can result in a significantly reduced estimation bias.
We first introduce the two-step estimation procedure as follows and then present the theoretical analysis.

Recall that in the first step, we obtain the WLSE $\wh \btheta^w$ by using the DNLSA algorithm.
Next, in the second step, we broadcast the WLSE to the local workers.
Then, we use $\wh{\btheta}^{w}$ as the initial value on the $k$th worker and perform one more step iteration to obtain a refined local estimator as follows,
	\begin{align}
		&\wh{\btheta}_k^{(2)} = \wh{\btheta}^{w} - \ddot{Q}_k^{-1}(\wh{\btheta}^{w}) \dot{Q}_k(\wh{\btheta}^{w}),~~~ \wh{\bSigma}_k^{(2)} = \ddot{Q}_k^{-1}(\wh{\btheta}^{w}) \label{hat_theta_k2}.
	\end{align}
	Then the local estimators $\wh{\btheta}_k^{(2)}$ and $\wh{\bSigma}_k^{(2)}$ are transmitted to the master, which consumes another round of communication. Thereafter, on the master, we obtain a TWLSE as
	\begin{align}
		\wh{\btheta}^{(2)} &= \Big\{\sum_{k = 1}^K \alpha_k \ddot{Q}_k(\wh{\btheta}^{w}) \Big\}^{-1} \Big\{\sum_{k = 1}^K \alpha_k \ddot{Q}_k (\wh{\btheta}^{w}) \wh{\btheta}_k^{(2)} \Big\} \label{hat_theta_2}\\
		& { \defeq} \Big\{\sum_{k = 1}^K \alpha_k \wh{\bSigma}_k^{(2) -1}  \Big\}^{-1} \Big\{\sum_{k = 1}^K \alpha_k \wh{\bSigma}_k^{(2) -1}  \wh{\btheta}_k^{(2)} \Big\}.\nonumber
	\end{align}
As one can see, the two-step estimator borrows the power of $\wh\btheta^w$
as a good initial estimator, which allows us to achieve lower estimation bias. We illustrate this point in the following theoretical analysis.


	\begin{theorem}[\textsc{Asymptotic Normality for TWLSE}] \label{thm.2}
		By assuming Conditions {(C\ref{con:noise})--(C\ref{con:converge})}, we have $\sqrt{N}(\wh \btheta^{(2)} - \btheta_0) =  (\wh\bSigma_2^w)^{-1} \{\bV(\btheta_0) + \bB_2(\btheta_0)\}$, with $\|\bB_2(\btheta_0)\| = O_p\{\sqrt{N}  (\log N)^{24}/n^2 \}$ and $\wh\bSigma_2^w = \sum_k \alpha_k \ddot{Q}_k(\wh\btheta^w) \to_p \bSigma_2$. Furthermore, we have $ \sqrt{N}(\wh \btheta^{(2)} - \btheta_0)\to_d N(0, \bSigma_2^{-1} \bSigma_1 \bSigma_2^{-1})$, under the condition that $n / \{N^{1/4} (\log N)^{12}\}\to \infty$.
	\end{theorem}

	The proof of Theorem \ref{thm.2} is provided in \ref{append.thm2}.
	In Theorem \ref{thm.2}, the asymptotic normality holds with local sample size $n/ \{N^{1/4} (\log N)^{12}\}\to \infty$, which allows for smaller local sample sizes than WLSE $\wh \btheta^w$.
In other words,
we see that the TWLSE trades off one more round of communication for a lower estimation bias.
In addition, this allows us to utilize more workers in the distributed system, which is particularly useful when more computing resources are accessible.
Following the same logic, we can refine the estimator multiple times to further reduce the asymptotic bias according to practical needs.

\subsection{Estimation Properties with Correlated and Heteroscedastic Error Terms}\label{corr_err}

In this section, we discuss the estimation properties for WLSE when the error terms are not i.i.d. distributed.
Specifically, we are interested in two cases.
The first is that $\ve_{i}$ is correlated across $1\le i\le N$ but is still identically distributed with $\var(\ve_{i}) = \sigma^2$.
The second is that we do not have a cross-sectional correlation for $\ve_i$ but heteroscedasticity arises such that $\var(\ve_{i}) = \sigma_i^2$.
For convenience, we let $\mE = \bSigma_e^{1/2} \wt\mE$ with $\wt \mE = (\wt \ve_i: i\in [N])^\top$, where $\wt \ve_i$s are i.i.d. random variables following a sub-Gaussian distribution with zero mean and unit variance.
Consequently, we can write $\cov(\mE) = \bSigma_e = (\sigma_{e,ij})$ and we then
discuss the detailed forms of $\bSigma_e$ in the next section.
Extensive numerical studies are provided in \ref{sigma_e_simulation} for Case I and \ref{gmm_simulation} for Case II.

\noindent
{\sc Case I: Correlated Error Terms.}

In this case, we should have $\diag(\bSigma_e)
 = \sigma^2\bI_N$, but $\bSigma_e$ has non-diagonal elements (i.e., $\sigma_{e,ij}\ne 0$ for some $1\le i,j\le N$).
 Intuitively, the WLSE can still be consistent when the cross-sectional correlation in $\bSigma_e$ is not strong.
 Particularly, we consider two structures for $\bSigma_e$ and study the estimation properties.
The first is a sparse structure for $\bSigma_e$.
We define $\mS_e = \{(i,j):\sigma_{e,ij}\ne 0, i\ne j\}$ as the index set for the
non-zero and non-diagonal elements in $\bSigma_e$.
A small $|\mS_e|$ implies a sparse structure for $\bSigma_e$.
The second is an equi-correlated structure for $\bSigma_e$.
Specifically, in this case we should have $\bSigma_e = \lambda\bI_N +
\gamma_N \one_N\one_N^\top$.
As a consequence, the error terms are equally correlated with $\cov(\ve_{i}, \ve_j) = \gamma_N$ for $i\ne j$.
We also study the estimation consistency of WLSE under the above two structures as follows.

\bep\label{prop:correlated_error}
Assume that Conditions (C1*), (C\ref{con:cov}), (C3*), (C\ref{con:param})--(C\ref{con:ident}), and (C7*)
 hold and that $K = O(N^{\zeta_1})$ with $\zeta_1 < 2 \delta$, where (C1*), (C3*)
 and (C7*) are given in \ref{append.tech_con} and $\delta$ is given in (C3*).
 Let $\wh \bSigma_{2e} = \sum_k \alpha_k \wh\bSigma_{2k,e}$ and $\wh \bSigma_{2k,e} = \ddot{Q}_k(\wh\btheta_k)$.
 Then, the following conclusions hold.\\
1. (Sparse $\bSigma_e$). Assume $\lambda_{\max}(\bSigma_e) \le c_0$ and $|\mS_e| = N^{\zeta_2}$ with $\zeta_1+\zeta_2<\delta$.
Then we have $\sqrt{N}(\wh\btheta^w - \btheta_0) = \wh\bSigma_{2e}^{-1} \{\bV_e(\btheta_{0}) + \bB_3(\btheta_0)\}$,
where $E\{\bV_e(\btheta_0)\} = \zero$, $\cov\{\bV_e(\btheta_0)\} = \bSigma_{1e}$ (defined in (C7*)),
$\|\bB_3(\btheta_0)\| = O_p\{K(\log N)^8 / \sqrt{N}  \} +O_p\{K |\mS_e| (\log N)^{6} / N^\delta\}$. \\
2. (Equi-correlated $\bSigma_e$). Assume $\gamma_N = O(N^{-\zeta_3})$ with $\zeta_3 >1/2$. Then we have $\sqrt{N} (\wh\btheta^w - \btheta_0) = \wh\bSigma_{2e}^{-1} \{\bV_e(\btheta_0) + \bB_4(\btheta_0)\}$, where $\|\bB_4(\btheta_0)\| = O_p\{K(\log N)^8 / \sqrt{N}  \} + O_p\{ \sqrt{N} (\log N)^4/ N^{\zeta_3} \}$.\\
In addition, we have $\wh \bSigma_{2e}\to_p \bSigma_{2e}$, where
$\bSigma_{2e}=\sum_k \alpha_k \bSigma_{2k,e}$ and $\bSigma_{2k,e} = E\{\ddot{Q}_k(\btheta_0)\}$.
\eep

The proof of Proposition \ref{prop:correlated_error} can be found in \ref{append.prop_err_cor}.
As implied by the results, the estimation bias can be controlled when the error terms are not seriously correlated.
Specifically, for the sparse case, we should have $K|\mS_e|/N^\delta\to 0$, which implies that the sparsity level (i.e., $|\mS_e|$) in $\bSigma_e$ should be controlled.
Compared to the diagonal $\bSigma_e$ case considered in Proposition \ref{prop.1}, the bias order is higher due to the extra cross-sectional dependence in $\bSigma_e$.
For the equi-correlated case, the cross-sectional dependence is controlled by the parameter $\gamma_N$, which should converge to zero as $N\to \infty$ to ensure an ignorable bias.
Subsequently, the asymptotic normality result can be readily obtained.


\noindent
{\sc Case II: Heteroscedastic Error Terms.}

In this case, we discuss the case in which $\bSigma_e = \diag\{\sigma_1^2,\sigma_2^2,\cdots, \sigma_N^2\}$ with non-identical variances $\sigma_i^2$.
Define $\ol\sigma^2 = N^{-1} \sum_i \sigma_i^2$ and $\ol \bSigma_e = \ol\sigma^2 \bI_N$.
Consequently, we can measure the distance from $\bSigma_e$ to the homoscedastic matrix $\ol \bSigma_e $ as $\|\bSigma_e - \ol \bSigma_e \| =
\max_i |\sigma_i^2 - \ol\sigma^2|\defeq \Delta$.
When $\Delta$ is small, the heteroscedasticity issue is not serious since
$\sigma_i^2$ are very close to each other.
Specifically, the consistency result can still hold when $\Delta$ is small, and we state the results rigorously in the following proposition.

\bep\label{prop.hetero_weak}
Assume Conditions (C1*), (C\ref{con:cov}), (C3*), (C\ref{con:param})--(C\ref{con:ident}), and (C7*). Further assume that $K = O(N^{\zeta_1}), \Delta = O\{N^{-\zeta_4}\}$ with $\zeta_1 < 2\delta, \zeta_4>1-\delta$ and $ \ol \sigma^2\le \tau_0$, where $\tau_0$ is a finite constant.
Then, we have $\sqrt{N}(\wh\btheta^w - \btheta_0) = \wh\bSigma_{2e}^{-1} \{\bV_e(\btheta_{0}) + \bB_5(\btheta_0)\}$,
where $\|\bB_5(\btheta_0)\| = O_p\{ K(\log N)^8 / \sqrt{N} \} + O_p\{ N^{1-\delta} \Delta (\log N)^6 \}$.
In addition, we have $E\{\bV_e(\btheta_{0})\} = \zero$, $\cov\{\bV_e(\btheta_{0})\} = \bSigma_{1e}$, and $\wh \bSigma_{2e}\to_p \bSigma_{2e}$.
\eep

As shown by Proposition \ref{prop.hetero_weak}, the consistency of WLSE can still be achieved when $\Delta$ is controlled, and the bias order could be ignored when $N$ goes to infinity, which can further lead to asymptotic normality.
However, in practice, we may still frequently encounter cases in which $\Delta$ is large \citep{anselin1988spatial,glaeser1996crime,lesage1999theory, lin2010gmm}.
In this case, a robust estimation framework is needed to obtain reliable estimation results.
It is recommended to employ the robust Generalized Method of Moments (GMM) estimation method proposed by \cite{lin2010gmm}.
Furthermore, with the GMM estimation framework, we can allow for potential endogeneity of the covariates $\bX$.
Our distributed estimation framework can be easily extended to the robust GMM (RGMM) method, and we provide the algorithm details in \ref{gmm_alg}.
Further numerical studies are also conducted to illustrate its robustness with the
heteroscedastic error terms and endogenous covariates.
The details are presented in \ref{gmm_simulation}.

Other than the correlated and heteroscedastic structures of $\bSigma_e$ discussed above, we can also assume specific forms for $\bSigma_e$.
For example, we may assume that $\bSigma_e$ depends on the exogenous covariates $\bX$. Specifically, we can follow \cite{zou2017covariance} to model $\bSigma_e$ as $\bSigma_e = \phi_0 \bI_N + \phi_1 \bA_1 + \cdots + \phi_M \bA_M$, where $\bA_m$ is the similarity matrix constructed from the $m$th covariate and $\phi_m$ is an unknown coefficient to be estimated.
We can also assume a spatial autoregression structure for $\mE$ \citep{das2003finite, lee2003best, kelejian2010specification},
i.e., $\mE = \rho_e\bW\mE + \e$.
This allows us to capture the spatial correlation pattern in $\mE$.
Since it might be beyond the scope of this work, we leave this as an interesting future research topic.

\section{DISTRIBUTED STATISTICAL INFERENCE}\label{sec:inference}


\subsection{Feasible Statistical Inference for WLSE and TWLSE}

Although the WLSE and TWLSE can conduct distributed estimation for the SAR model, they still cannot allow for distributed statistical inference simultaneously.
For convenience, in the following, we assume that $\ve_i$ follows a normal distribution with covariance $\sigma_\ve^2$.
Note that in Theorem \ref{thm.1}, the asymptotic covariance takes the form $\bSigma_2^{-1}\bSigma_1\bSigma_2^{-1}$.
Specifically, we have $\bSigma_2 = \sum_k \alpha_k \bSigma_{2k}$, where $\bSigma_{2k} = E\{\ddot{Q}_k (\btheta_{0})\}$. We can estimate
$\bSigma_{2k}$ on each worker simply by $\wh \bSigma_{2k} = \ddot Q_k(\wh \btheta_k)$ (for the WLSE)
or $\wh \bSigma_{2k} = \ddot Q_k(\wh \btheta_k^{(2)})$ (for the TWLSE).
However, the estimation for $\bSigma_1$ is more challenging.
More specifically, we have $\bSigma_1 = \sum_{k,l = 1}^K \sqrt{\alpha_k\alpha_l}\bSigma_{1kl}$,
where $\bSigma_{1kl} = { \sqrt{N_kN_l}}\cov\{\dot{Q}_k(\btheta_0), \dot{Q}_l(\btheta_0)\}$ takes the form,
\begin{align}
	&{\bSigma}_{1kl, \rho} = \frac{4}{\sqrt{N_k N_l}}  [\sigma_\ve^4 \{\tr( \bXi_{k} \bXi_{l}^\top) + \tr( \bXi_{k}\bXi_{l})\} 
	+ \sigma_\ve^2 \bU_{1k} \bU_{1l}^\top], \nonumber\\
	& {\bSigma}_{1kl, \rho\bbeta} = \frac{ 4 \sigma_\ve^2}{\sqrt{N_k N_l}} (\bU_{1k}\bU_{2l}^\top), ~~~ {\bSigma}_{1kl, \bbeta} = \frac{ 4 \sigma_\ve^2}{\sqrt{N_k N_l}} (\bU_{2k} \bU_{2l}^\top),\label{true_Sigma_1}
\end{align}
where 
$\bXi_{k} \in \mR^{N \times N}$,
$\bU_{1k} \in \mR^{1 \times N}$ and $\bU_{2k} \in \mR^{p \times N}$ and
the specific forms
are discussed in detail in \ref{asymp_cov}.
Through careful investigation of (\ref{true_Sigma_1}), we find that it involves
a typical term, $\tr\{\bM (\bS_0^\top \bS_0)^{-1}\}$, where
$\bM\in \mR^{N\times N}$
is a given matrix and $\bS_0 = \bI - \rho_0\bW$.
Generally speaking, the computation is difficult since it requires computing the inverse of a high-dimensional matrix $\bS_0^\top \bS_0$.
To this end, we borrow the idea from \cite{huang2019least} to estimate the value $\tr\{\bM (\bS_0^\top \bS_0)^{-1}\}$ using the sample data instead.

%
%
%

Specifically, we note that
$E_{\bX, \ve}(\by^\top \bM \by) =\wt\sigma^2 \tr\{ \bM (\bS_0^\top \bS_0)^{-1}\}$, $ (1-\sigma_\ve^2/\wt\sigma^2) E_\ve (\by^\top \bM \by) = (\bS_0^{-1} \bX \bbeta_0)^\top \bM (\bS_0^{-1} \bX \bbeta_0)$ and $ E_\ve (\by) = \bS_0^{-1} \bX \bbeta_0$,
where
$ E_{\bX, \ve}(\cdot)$ denotes expectation on $(\bX, \ve)$, $E_{\ve}(\cdot)$ denotes expectation on $\mE$, and $\wt \sigma^2 = \bbeta_{0}^\top \bSigma_X\bbeta_{0} + \sigma_\ve^2$.
Here, we treat $\bX_i $ as independent and identically distributed random variables with mean $ \zero$
and covariance $\bSigma_X$ for convenience.
Consequently, $ \wh { \sigma}^{-2} (\by^\top \bM\by)$ and $(1- \wh\sigma_\ve^2/\wh{\sigma}^2) \by^\top\bM\by$ could serve as estimators for $ \tr\{\bM (\bS_0^\top \bS_0)^{-1}\} $ and $(\bS_0^{-1} \bX \bbeta_0)^\top \bM (\bS_0^{-1} \bX \bbeta_0)$ respectively, where $ \wh {\sigma}^{2}, \wh\sigma_\ve^2$ are sample estimates for $\wt\sigma^2$ and $\sigma_\ve^2$ respectively.
By exploiting this property, we can extend the covariance estimation of \cite{huang2019least} to our case with covariates $\bX$ and obtain the following estimator $\wh\bSigma_{1kl}$,
\begin{align}
	\wh{\bSigma}_{1kl, \rho} &= \frac{4}{\sqrt{N_k N_l}} \big[\wh\sigma_\ve^4 \big\{ \tr(\bXi_k^\dag \bXi_l^\dag) +\tr(\bV_{1k}^\top \bV_{2l}) + \wh\sigma^{-2} (\bT_{1k} \bT_{2l}^\top + \bT_{2k} \bT_{1l}^\top )\big\} +  \wh\sigma_\ve^2 \bT_{1k} \bT_{1l}^\top\big] \nonumber\\
	\wh{\bSigma}_{1kl, \rho\bbeta} &= - \frac{4 \wh\sigma_\ve^2}{\sqrt{N_k N_l}} \bT_{1k} \bT_{3l}^\top, ~~~ \wh{\bSigma}_{1kl, \bbeta} =  \frac{4 \wh\sigma_\ve^2}{\sqrt{N_k N_l}}  \bT_{3k} \bT_{3l}^\top. \label{est_Sigma_1}
\end{align}
The quantities $\bXi_k^\dag$, $\bV_{1k}, \bV_{2k}, \bT_{1k}, \bT_{2k}$ and $\bT_{3k}$ are calculated as follows.
Define $\dot \bD_\rho = \partial \bD / \partial \rho = -2 \rho \mathbf{D}^2 \operatorname{diag}(\mathbf{W}^{\top} \mathbf{W})$, $\mathbf{J}_k=\sum_{i \in \mathcal{S}_k} \mathbf{e}_i \mathbf{e}_i^{\top} \in \mathbb{R}^{N \times N}$. Then, we have
\begin{align}
	& \bXi_k^\dag =  (\bS^\top \bS \dot{\bD}_\rho - \bS^\top \bW \bD - \bW^\top \bS \bD) \bJ_k  \bD \in \mR^{N \times N} ,\nonumber\\
	& \bV_{1k} =  \bD \bS^\top  \bJ_k \in \mR^{N \times N}, ~~~ \bV_{2l} = \wt\bM \bJ_l  \bD \bS^\top \mR^{N \times N}, \nonumber\\
	& \wt\bM =  \dot{\bD}_\rho \bS^\top \bS \dot{\bD}_\rho -  \dot{\bD}_\rho \bS^\top \bW \bD - \dot{\bD}_\rho \bW^\top \bS \bD - \bD \bW^\top \bS \dot{\bD}_\rho + \bD \bW^\top \bW \bD - \bD \bS^\top \bW \dot{\bD}_\rho \mR^{N \times N}, \nonumber\\
	& \bT_{1k}  = \by^\top \bW^\top \bS \bD  \bJ_k   \bD \bS^\top \in \mR^{1 \times N} , \nonumber\\
	& \bT_{2k} = \by^\top \bS^\top \bW \bD \bJ_k  \bD \bS^\top  \in \mR^{1\times N}, ~~~ \bT_{3k} = \bX^\top \bS \bD \bJ_k \bD \bS^\top \in \mR^{p \times N} .\label{eq:Xi_ks}
\end{align}
by replacing
$\btheta$ in the above formulation with $\wh \btheta.$

Although the forms in \eqref{est_Sigma_1} are slightly complicated,
one should note that it does not involve the
inverse of a high-dimensional matrix; therefore, it is more computationally tractable.
Next, we establish the following theorem that the covariance estimator $\wh \bSigma_1 =  \sum_{k,l = 1}^K \sqrt{\alpha_k\alpha_l}\wh \bSigma_{1kl}$ provides a consistent estimation of $\bSigma_1$.
This extends the consistency result of the covariance estimator proposed by \cite{huang2019least} to the SAR model with exogenous covariates information.

\bet[\textsc{Consistency for $\wh\bSigma_{1}$}]\label{thm3:est_sigma1_consistency}
Under Conditions (C\ref{con:noise}) and (C\ref{con:network}), we have $\wh\bSigma_{1} \to_p \bSigma_{1}$ as $N \to \infty$.
\eet

The proof of Theorem \ref{thm3:est_sigma1_consistency} is provided in \ref{append:thm3}.
It is noteworthy that although $\wh \bSigma_{1kl}$ is computationally feasible, it is not communicationally efficient
for a distributed system since it utilizes the data from the $k$th and $l$th worker.
Specifically, it requires transmitting a set of $N\times N$ dimensional matrices (e.g., $\bXi_k^\dag, \bV_{1k}, \bV_{2k}$) from the workers to the master to calculate the estimator in \eqref{est_Sigma_1}.
Therefore, we further discuss how to
conduct a valid statistical inference with low communication cost in a distributed system
in the subsequent section.

\begin{remark}
We remark that the calculation of matrices and vectors in \eqref{eq:Xi_ks}
still requires local network information instead of the full data information.
We use $\bXi_k^\dag$ for illustration.
Note that we can write $\bXi_k^\dag = \bXi_{k,1}^\dag\bXi_{k,2}^{\dag\top}$,
where $\bXi_{k,1}^\dag =  (\bS^\top \bS \bD_\rho - \bS^\top \bW \bD - \bW^\top \bS \bD) \bJ_k^{(\cdot, \mS_k)} \in \mR^{N\times N_k}$ and $\bXi_{k,2}^\dag = \bD \bJ_k^{(\cdot, \mS_k)}
\in \mR^{N\times N_k}$.
Here, recall that $\bJ_k^{(\cdot, \mS_k)}$ is a sub-matrix of $\bJ_k$ with column indices in $\mS_k$.
According to the formulation of $\bXi_{k,1}^\dag$ and $\bXi_{k,2}^\dag$, we observe that it needs the information of node $j$ if it is connected to nodes $i\in \mS_k$ by up to a second-order network connection, which is stated in Section \ref{subsec:estimation}.
As a consequence, the calculation of $\bXi_{k}^\dag$ only requires local-network information despite it being of dimension $N\times N$.
For inference purpose, we need to calculate
$\sum_{k,l} \tr(\bXi_k^\dag\bXi_l^\dag)$ on the master as shown in \eqref{est_Sigma_1}.
This requires communicating the matrix $\bXi_k^\dag\in \mR^{N\times N}$ from the workers to the master, which may incur a high communication cost for the distributed system.
\end{remark}

\subsection{Communicationally Efficient Statistical Inference}\label{subsec:randon_proj}

In this section, we discuss how to estimate the asymptotic covariance, i.e., $\bSigma_2^{-1}\bSigma_1\bSigma_2^{-1}$ in a distributed system.
First, to estimate $\bSigma_2 = \sum_k \alpha_k \bSigma_{2k}$ on the master, it is sufficient to transmit the estimator $\wh \bSigma_{2k}$ from the $k$th worker to the master.
However, estimating $\bSigma_1 = \sum_{k,l}\sqrt{\alpha_k\alpha_l}\bSigma_{1,kl}$ is more complicated, because to calculate $\wh\bSigma_{1,kl}$ by \eqref{est_Sigma_1} on the master, one needs to obtain several matrices as $ \bXi_k^\dag\in \mR^{N\times N}$ from the $k$th worker.
Particularly, we note that the dimension of $ \bXi_k^\dag$ is $N\times N$, which implies that transmitting the matrix from the workers to the master will require high communication costs especially when $N$ is large.

To reduce the communication cost, we consider a random projection method motivated by the JL Lemma \citep{johnson1984extensions},
which states that the distance between two vectors can be preserved after
projecting them into a low-dimensional space with random matrices.
The idea has been widely used in recent machine learning literature \citep{bingham2001random, becchetti2019oblivious, meister2019tight}.
Therefore, this motivates us to project the high-dimensional terms in \eqref{est_Sigma_1} into low-dimensions using a similar technique, which could improve the communication efficiency.
Specifically, on each worker, we generate random matrices
$\bR_1, \bR_2 \in \mR^{d \times N}$ with $d \ll N$.
The entries of $\bR_1, \bR_2$  are independently generated
from $N(0,1/d)$, and consequently it holds that $E(\bR_m^\top\bR_m) = \bI_N$ for $m = 1,2$.
Instead of directly transmitting the matrices as $ \bXi_k^\dag$ from each worker to the master, we project the estimators to lower dimensions using $\bR_1, \bR_2$.
Specifically, the projected version of the corresponding matrices (vectors) is defined as follows,
\begin{alignat}{4}
	&  \bXi_{k,1}^{\dag \text{R}} \defeq \bR_1 \bXi_k^{\dag} \bR_2^\top  \in \mR^{d \times d}, && \bXi_{k,2}^{\dag \text{R}} \defeq \bR_2 \bXi_k^{\dag} \bR_1^\top  \in \mR^{d \times d},  \bT_{3k}^{\text{R}} \defeq \bT_{3k} \bR_1^\top  \in \mR^{p \times d} \nonumber\\
	& \bV_{mk}^{\text{R}} \defeq \bR_1 \bV_{mk} \bR_2^\top  \in \mR^{d \times d}  , && \bT_{mk}^{\text{R}} \defeq \bT_{mk} \bR_1^\top  \in \mR^{1 \times d} \ (m = 1,2). \label{eq:matrix_R}
\end{alignat}
Through the above, we could project all terms in \eqref{est_Sigma_1} into a low-dimensional space.
Transmitting the above matrices will largely reduce the communication cost with small $d$.
We explain the basic ideas about the above random projection method as follows.
As stated by the JL Lemma \citep{johnson1984extensions}, for a non-random vector $\bv\in \mR^N$ with $\|\bv\|^2  =1$, one could project it to a low-dimensional vector as
$\bv_1 = \bR_1\bv\in \mR^d$ with a random projection matrix $\bR_1$ (as defined above),
and it holds that $\|\bv_1\|^2 - \|\bv\|^2 \to_p 0$ as $N\to\infty$ as long as
$d \gtrsim  \log N$.
Motivated by this fact, we utilize the property of random projection and aim to show that the values in \eqref{est_Sigma_1} can be approximated by using the projected matrices/vectors in \eqref{eq:matrix_R}.
Using $\tr(\bXi_{k,1}^{\dag \text{R}}  \bXi_{l,2}^{\dag \text{R}})$ as an example, we can show that
\begin{align*}
E\left\{\tr(\bXi_{k,1}^{\dag \text{R}}  \bXi_{l,2}^{\dag \text{R}} ) | \mD\right\}  = \tr(\bXi_{k,1}^{\dag}  \bXi_{l,2}^{\dag} ),
\end{align*}
where $\mD = \{\by,\bX\}$. Note that $\tr(\bXi_{k,1}^{\dag}  \bXi_{l,2}^{\dag} )$ is used for calculating $\wh \bSigma_{1kl,\rho}$ in \eqref{est_Sigma_1}.
In the following we basically show that
$N^{-1} |\tr(\bXi_{1}^{\dag \text{R}}  \bXi_{2}^{\dag \text{R}} )-\tr(\bXi_{1}^{\dag}  \bXi_{2}^{\dag} )| = o_p(1)$, where $\bXi_{1}^{\dag \text{R}} = \sum_k \alpha_k \bXi_{k,1}^{\dag \text{R}}$ and $\bXi_{2}^{\dag \text{R}} = \sum_l \alpha_l \bXi_{l,2}^{\dag \text{R}}$.
The same convergence properties can be demonstrated for other terms in \eqref{eq:matrix_R} as well.
As a result, the low-dimensional matrices in \eqref{eq:matrix_R} are substitutes
for their counterparts in \eqref{est_Sigma_1} and we could prove that the difference can be ignored with high probability.

In practice, to ease the computation, one can generate matrices $\bR_1$ and $\bR_2$ as sparse matrices using the package \href{https://scikit-learn.org/stable/modules/random_projection.html#sparse-random-projection}``{\sf scikit-learn}''.
This could make the projection matrix sparse, and thus, it is easy to calculate the amounts in \eqref{eq:matrix_R}.
Accordingly, the random projected estimator $\wh \bSigma_{1kl}^\tR$
is given by
\begin{align}
	& \wh{\bSigma}_{1kl, \rho}^{\text{R}} = \frac{4}{\sqrt{N_k N_l}} \Big\{  \wh\sigma_\ve^4  \big\{  \tr(\bXi_{k,1}^{\dag \text{R}}  \bXi_{l,2}^{\dag \text{R}} ) +\tr(\bV_{1k}^{\text{R}\top } \bV_{2l}^{\text{R}}) + \wh\sigma^{ -2} (\bT_{1k}^{\text{R}} \bT_{2l}^{\text{R}\top } + \bT_{2k}^{\text{R}}  \bT_{1l}^{\text{R}\top } )\big\}  + \wh\sigma_\ve^2  \bT_{1k}^{\text{R}}  \bT_{1l}^{\text{R}\top } \Big\}, \nonumber\\
	& \wh{\bSigma}_{1kl, \rho\bbeta}^{\text{R}}  = - \frac{4 \wh\sigma_\ve^2}{\sqrt{N_k N_l}} \bT_{1k}^{\text{R}}  \bT_{3l}^{\text{R}\top } , ~~~ \wh{\bSigma}_{1kl, \bbeta}^{\text{R}} =  \frac{4 \wh\sigma_\ve^2}{\sqrt{N_k N_l}}  \bT_{3k}^{\text{R}}  \bT_{3l}^{\text{R}\top }\label{Sigma1_proj_k}
\end{align}
Here, we remark that the matrices $\bR_1$ and $\bR_2$ should remain the same for all workers by setting the same random seed in implementation, and the estimates could be obtained by substituting $\btheta$ with $\wh\btheta$.

One can easily verify that $E(\wh \bSigma_{1kl}^\tR| {\mD}) = \wh \bSigma_{1kl}$.
Intuitively, $\wh \bSigma_{1kl}^\tR$ can play a role as an approximation of
$\wh \bSigma_{1kl}$.
Since $\wh \bSigma_{1}$ is a consistent estimator for $\bSigma_{1}$, as implied by Theorem \ref{thm3:est_sigma1_consistency}, it remains to be verified that
$\wh \bSigma_{1}^\tR = \sum_{k,l}\sqrt{\alpha_k\alpha_l}\wh \bSigma_{1kl}^\tR$
can serve as a good approximation of
$\wh\bSigma_{1}$ under certain conditions.
In the following, we establish the consistency result of our random projection estimator.

\begin{theorem}\label{thm4:random_proj}
Assume Conditions (C\ref{con:noise})--(C\ref{con:converge}) and  $d \gtrsim \log N$.
Then, we have $\wh{\bSigma}_{1}^{\textup{R}} \to_p \bSigma_{1}$ as $N\rightarrow\infty$.
\end{theorem}

The proof of Theorem \ref{thm4:random_proj} is provided in \ref{append.thm4}.
According to Theorem \ref{thm4:random_proj}, the random projection estimator
is consistent as long as we have $d\gtrsim \log N$.
This result is in agreement with the classical Johnson-Lindenstrauss Lemma \citep{SanjoyDasgupta2003AnEP}.
The situation in our case is slightly different due to the complex expression of the matrices and vectors we need to project.
More importantly, the communication cost greatly decreases from $O(N^2)$ to $O(d^2) = O\{(\log N)^2\}$ after the random projection procedure.
Then, the asymptotic covariance can be estimated on the master
with $\wh \bSigma_2^{-1}\wh \bSigma_1^\tR\wh \bSigma_2^{-1}$.
We summarize the distributed estimation and corresponding inference procedures in Algorithm \ref{alg:dsar}.

\begin{algorithm}
	\caption{Distributed Estimation and Inference for the SAR Model}
	\begin{algorithmic}[1]
		\item {\bf Step 1.} On each worker $k = 1, \cdots, K$, minimize $Q_k(\btheta)$ to obtain $\wh{\btheta}_k$. 
		\item {\bf Step 2.} Then transmit $\wh{\btheta}_k$ and $\ddot{Q}_k(\wh\btheta_k)$ to the master.
		\item {\bf Step 3.} Calculate WLSE $\wh{\btheta}^w$ according to \eqref{wlse} on the master.
		\item {\bf Step 4.} Broadcast $\wh\btheta^w$ to the workers.
		\item {\bf Step 5.} On each worker $k = 1, \cdots, K$: use $\wh{\btheta}^w$ to perform a one-step iteration to obtain a refined local estimator $\wh{\btheta}_k^{(2)}$ by \eqref{hat_theta_k2}. Calculate $\wh\bXi_{k,1}^{\dag \text{R}}, \wh\bXi_{k,2}^{\dag \text{R}}, \wh\bV_{1k}^{\tR}, \wh\bV_{2k}^\tR, \wh\bT_{1k}^\tR, \wh\bT_{2k}^\tR, \wh\bT_{3k}^\tR$ using $\wh{\btheta}_k^{(2)}$ by \eqref{eq:matrix_R}.
		\item {\bf Step 6.} Transmit $\wh{\btheta}_k^{(2)}$, $\wh{\bSigma}_{k}^{(2)}$ and
		$\wh\bXi_{k,1}^{\dag \text{R}}, \wh\bXi_{k,2}^{\dag \text{R}}, \wh\bV_{1k}^{\tR}, \wh\bV_{2k}^\tR, \wh\bT_{1k}^\tR, \wh\bT_{2k}^\tR, \wh\bT_{3k}^\tR$ to the master.
		\item {\bf Step 7.} Calculate TWLSE $\wh{\btheta}^{(2)}$ by \eqref{hat_theta_2} and
		$\wh \bSigma_{2} = \sum_k \alpha_k \wh \bSigma_{2k}$, ${\wh \bSigma_1^\tR} = \sum_{k,l}\sqrt{\alpha_k\alpha_l} \wh \bSigma_{1kl}^{\tR}$ on the master.
%
		\State {\bf Output:} Estimators WLSE and TWLSE, and the corresponding estimated asymptotic covariance $\wh{\bSigma}_{2}^{-1} \wh{\bSigma}_{1}^\tR \wh{\bSigma}_{2}^{-1}$.
	\end{algorithmic}\label{alg:dsar}
\end{algorithm}

\section{NUMERICAL STUDIES}\label{sec:num}

\subsection{Simulation Models and Settings}\label{subsec:simulation_setting}

To demonstrate the finite sample performance of the DNLSA algorithm, we conduct a number of simulation studies in this section.
Given the network size $N$, we first generate the adjacency matrix $\bA = (a_{ij}) \in \mR^{N \times N}$. Note that $\bA$ is not necessarily symmetric.
Specifically, we generate two types of networks as follows.


\textbf{Example 1.} (Stochastic Block Model) We first consider the stochastic block model \citep{wang1987stochastic,nowicki2001estimation} for generating the network.
The SBM assumes that nodes within the same block are more likely to be connected than nodes from different blocks. We set $M=20$ blocks and follow \cite{nowicki2001estimation} to randomly assign each node a latent label $k \in \{1,2,\cdots, M\}$ with equal probability $1/M$. Next, let $P(a_{ij} = 1) = 20 N^{-1}$ if $i$ and $j$ are in the same block, and $P(a_{ij}=1) = 2N^{-1}$ otherwise.

\textbf{Example 2.} (Power-Law Distribution) We follow \cite{clauset2009power} to
generate a network whose nodes' in-degrees follow the
the power-law distribution. Specifically, for each node $i$, we first generate its in-degree $d_{i}=\sum_{j} a_{j i}$ according to the discrete power-law distribution with $P\left(d_{i}=k\right)=c k^{-\alpha}$, where $c$ is a normalizing constant and the parameter is set as $\alpha = 3$.
Then, we randomly select $d_i$ nodes as the potential followers of node $i$. This setting could guarantee that the majority of nodes have few edges but a small number of nodes (e.g., influential people) have a large number of edges \citep{barabasi1999emergence}.
As a consequence, it can reflect the ``superstar effect'' in networks.


Next, for each example, we generate the covariates
 $X_{ij} (1 \le i \le N, 1 \le j \le p)$ from the standard normal distribution $N(0,1)$ independently with $p = 5$.
The error term $\ve_i$ ($1\le i\le N$) is i.i.d. generated from standard normal distribution $N(0,1)$.
 We also conduct a simulation study when $\ve_i$ follows the $t$-distribution, and the details are given in \ref{random_t_dist}.
The true parameters of the SAR model are set as,
$\rho=0.4,\ \beta_1=0.2,\ \beta_2=0.4,\ \beta_3=0.6,\ \beta_4=0.8, \ \text{and} \ \beta_5=1.0,$
which remain the same across the two examples.
We set the sample size and number of workers as $N \in \{2,4,10,20\}\times 10^3$ and $K \in \{10,20,40\}$, respectively.
In addition, the local sample size on the $k$th worker is specified as
$N_k = N/K$, if $N$ can be divided exactly by $K$.
Otherwise, we first distribute $[N/K]$ nodes on each worker and then uniformly
distribute the remaining nodes on all workers,
where $[r]$ denotes the integer part of $r$.

For comparison, we implement the OS estimator \citep{zhang2013communication, battey2015distributed, XiangyuChang2017DivideAC}, one-step estimator (WLSE) as well as the two-step estimator (TWLSE) for a distributed estimation.
Specifically, the OS estimator is obtained by taking the average of the local estimators of all workers as
$
\wh \btheta^{\os} = K^{-1}\sum_k \wh \btheta_k.
$
In the following section, we introduce how we measure the performance under the above model settings and evaluate the finite sample performance.

%

\subsection{Performance Measurements and Simulation Results}\label{subsec:comp}


To ensure a reliable evaluation,
the experiment is repeated for a total of $R = 500$ times under each model setting.
For the $r$th replicate, denote the estimator as
$\wh{\btheta}^{(r)} = (\wh \theta_j^{(r)})^\top$. The corresponding global estimator is recorded as $\wt{\btheta}^{(r)} = (\wt \theta_j^{(r)})^\top$, which is estimated by using the whole data information.
Then, the root mean square error (RMSE) is calculated for the $j$th parameter estimator as $\text{RMSE}_{\wh{\theta}_j} = \{R^{-1} \sum_r (\wh{\theta}_j^{(r)}	-  \theta_{0,j})^2 \}^{1/2}$. Similarly, the RMSE for the global estimator is expressed as $\text{RMSE}_{\wt{\theta}_j} = \{R^{-1} \sum_r (\wt{\theta}_j^{(r)}	- \theta_{0,j})^2 \}^{1/2}$.
To evaluate the estimation efficiency,
we define the relative estimation efficiency (REE) with respect to each estimator as $\text{REE}_j = \text{RMSE}_{\wt{\theta}_j} / \text{RMSE}_{\wh{\theta}_j}$.
Consequently, the estimator attains global efficiency if
the REE is close to 1.
Next, we evaluate the performance of the statistical inference.
For the $j$th parameter, the $95\%$ confidence interval is constructed as $\text{CI}_j^{(r)} = (\wh{\theta}_j^{(r)} - z_{0.975} \wh{\text{SE}}_j^{(r)}, \wh{\theta}_j^{(r)} + z_{0.975} \wh{\text{SE}}_j^{(r)})$,
where $\wh{\text{SE}}_j^{(r)}$ is the estimation of the standard error obtained from the $j$th diagonal element of $\wh\bSigma_{2}^{-1} \wh\bSigma_{1}^{\text{R}} \wh\bSigma_{2}^{-1}$ given in Algorithm \ref{alg:dsar}, and
$z_\alpha$ is the $\alpha$ quantile of the standard normal distribution.
Here, we set $ d = [\log N]+1$ when calculating $ \wh\bSigma_{1}^{\text{R}}$ in \eqref{Sigma1_proj_k},
where $[\cdot]$ denotes the integer part.
Then the coverage probability (CP) of the $j$th parameter estimation is calculated as $\text{CP}_{j}=R^{-1} \sum_{r=1}^{R} I \{\theta_{0,j} \in \text{CI}_{j}^{(r)}\}$.
We remark that the CP is not calculated for the OS estimator since the corresponding Hessian matrix is not transmitted from workers to the master for this method.

The simulation results can be found in Table \ref{table:example1}--\ref{table:example2}.
Similar patterns are observed for both the SBM and power-law distribution networks.
First, one could observe that under the same setting of worker number $K$,
the REEs of both the OS and WLSE show an increasing trend as $N$ increases.
Specifically, we take the estimator $\wh\rho$ of the SBM network for example. With $K =40$,
the REEs of the WLSE are approximately 0.774 when $N=2000$ and can reach 0.962 when $N = 20000$, which is in line with the results in Theorem \ref{thm.1}.
Next, the REEs of the TWLSE of the SBM network achieve global efficiency in nearly all $N$ and $K$ settings.
In the power-law distribution network, the REEs of the TWLSE show a similar increasing trend as those of the OS and WLSE, and it attains the global efficiency with $\text{REE} \approx 1$
 as the sample sizes increase.
In summary, the proposed WLSE and TWLSE are obviously more efficient than the OS estimator across all settings, and the TWLSE can perform much better than the WLSE,  which corroborates Theorem \ref{thm.2} very well.
Moreover, the TWLSE method exhibits better performance with a large $K$, in which case smaller local sample sizes are allowed and the estimation accuracy is still preserved.
Last, we observe that the CPs for both the WLSE and TWLSE methods are all around 95\%
for a large $N$, which indicates the validity of our proposed statistical inference procedure.

\begin{figure}[htpb!]
	\begin{center}
		\includegraphics[width=0.5\textwidth]{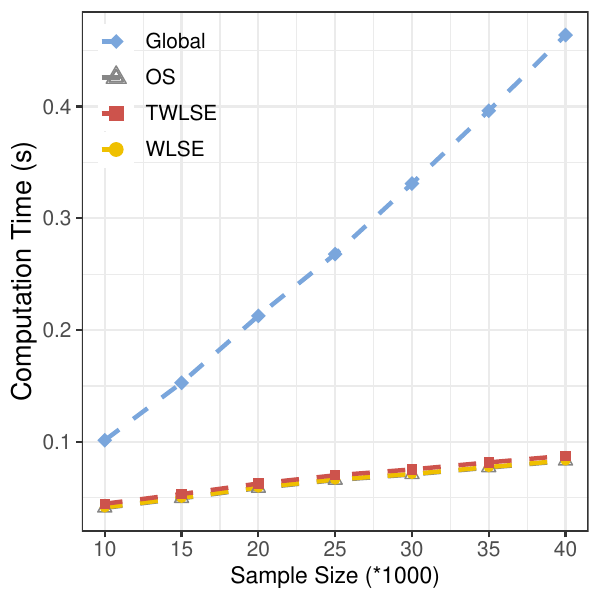}
		\caption{\small Computational time (in seconds) for different estimators. The global estimator, OS estimator, WLSE and TWLSE are shown in blue, gray, yellow and red, respectively.}
		\label{fig:compute_time}
	\end{center}
\end{figure}

Next, we illustrate the computational efficiency of our proposed methods.
We compare the time cost of distributed algorithms with the global estimation.
To this end, we use a machine containing
	18 CPU cores and 384 GB of RAM.
We use a single CPU core for the global estimation and all CPU cores for the distributed estimators with the Spark system.
	We fix $K = 36$ and increase $N$ from 10000 to 40000; the computational time is shown in Figure \ref{fig:compute_time}.
One could observe that the global estimation requires a much higher computational cost than the distributed estimators, especially when $N$ is large.
In addition, both the OS and WLSE require a lower computational cost than the TWLSE, which is as expected since lower communication and local computation costs are consumed.


\begin{sidewaystable}[thp]\footnotesize
	\centering
	\caption{``REE" results for Example 1 with 500 replications. The performances are evaluated for different sample sizes $N(\times10^3)$ and numbers of workers $K$. The corresponding CPs are displayed in parentheses.}\label{table:example1}
	\scalebox{0.9}{
\begin{tabular}{c|c|ccccccc|c|c|ccccccc}
	\hline
	$N$ &
	$K$ &
	Estimation &
	$\rho$ &
	$\beta_1$ &
	$\beta_2$ &
	$\beta_3$ &
	$\beta_4$ &
	$\beta_5$ &
	$N$ &
	$K$ &
	Estimation &
	$\rho$ &
	$\beta_1$ &
	$\beta_2$ &
	$\beta_3$ &
	$\beta_4$ &
	$\beta_5$ \\ \hline
	\multirow{9}{*}{2} &
	\multirow{3}{*}{10} &
	OS &
	0.967 &
	0.977 &
	0.985 &
	0.978 &
	0.985 &
	0.987 &
	\multirow{9}{*}{10} &
	\multirow{3}{*}{10} &
	OS &
	0.996 &
	0.997 &
	0.999 &
	1.002 &
	0.995 &
	0.999 \\
	&
	&
	WLSE &
	\begin{tabular}[c]{@{}c@{}}0.979\\ (0.952)\end{tabular} &
	\begin{tabular}[c]{@{}c@{}}1.002\\ (0.914)\end{tabular} &
	\begin{tabular}[c]{@{}c@{}}1.004\\ (0.922)\end{tabular} &
	\begin{tabular}[c]{@{}c@{}}0.999\\ (0.924)\end{tabular} &
	\begin{tabular}[c]{@{}c@{}}1.000\\ (0.940)\end{tabular} &
	\begin{tabular}[c]{@{}c@{}}1.002\\ (0.938)\end{tabular} &
	&
	&
	WLSE &
	\begin{tabular}[c]{@{}c@{}}0.995\\ (0.948)\end{tabular} &
	\begin{tabular}[c]{@{}c@{}}1.000\\ (0.918)\end{tabular} &
	\begin{tabular}[c]{@{}c@{}}1.000\\ (0.930)\end{tabular} &
	\begin{tabular}[c]{@{}c@{}}1.001\\ (0.912)\end{tabular} &
	\begin{tabular}[c]{@{}c@{}}1.000\\ (0.916)\end{tabular} &
	\begin{tabular}[c]{@{}c@{}}0.999\\ (0.934)\end{tabular} \\
	&
	&
	TWLSE &
	\begin{tabular}[c]{@{}c@{}}1.000\\ (0.958)\end{tabular} &
	\begin{tabular}[c]{@{}c@{}}1.000\\ (0.908)\end{tabular} &
	\begin{tabular}[c]{@{}c@{}}1.000\\ (0.922)\end{tabular} &
	\begin{tabular}[c]{@{}c@{}}1.000\\ (0.924)\end{tabular} &
	\begin{tabular}[c]{@{}c@{}}1.000\\ (0.934)\end{tabular} &
	\begin{tabular}[c]{@{}c@{}}1.000\\ (0.938)\end{tabular} &
	&
	&
	TWLSE &
	\begin{tabular}[c]{@{}c@{}}1.000\\ (0.954)\end{tabular} &
	\begin{tabular}[c]{@{}c@{}}1.000\\ (0.920)\end{tabular} &
	\begin{tabular}[c]{@{}c@{}}1.000\\ (0.928)\end{tabular} &
	\begin{tabular}[c]{@{}c@{}}1.000\\ (0.912)\end{tabular} &
	\begin{tabular}[c]{@{}c@{}}1.000\\ (0.912)\end{tabular} &
	\begin{tabular}[c]{@{}c@{}}1.000\\ (0.934)\end{tabular} \\ \cline{2-9} \cline{11-18}
	&
	\multirow{3}{*}{20} &
	OS &
	0.275 &
	0.850 &
	0.849 &
	0.767 &
	0.873 &
	0.828 &
	&
	\multirow{3}{*}{20} &
	OS &
	0.984 &
	0.991 &
	0.989 &
	1.005 &
	0.996 &
	1.001 \\
	&
	&
	WLSE &
	\begin{tabular}[c]{@{}c@{}}0.876\\ (0.926)\end{tabular} &
	\begin{tabular}[c]{@{}c@{}}0.996\\ (0.908)\end{tabular} &
	\begin{tabular}[c]{@{}c@{}}1.000\\ (0.924)\end{tabular} &
	\begin{tabular}[c]{@{}c@{}}0.999\\ (0.928)\end{tabular} &
	\begin{tabular}[c]{@{}c@{}}0.994\\ (0.932)\end{tabular} &
	\begin{tabular}[c]{@{}c@{}}0.988\\ (0.940)\end{tabular} &
	&
	&
	WLSE &
	\begin{tabular}[c]{@{}c@{}}0.975\\ (0.942)\end{tabular} &
	\begin{tabular}[c]{@{}c@{}}1.001\\ (0.918)\end{tabular} &
	\begin{tabular}[c]{@{}c@{}}1.000\\ (0.928)\end{tabular} &
	\begin{tabular}[c]{@{}c@{}}1.001\\ (0.912)\end{tabular} &
	\begin{tabular}[c]{@{}c@{}}0.999\\ (0.904)\end{tabular} &
	\begin{tabular}[c]{@{}c@{}}0.997\\ (0.934)\end{tabular} \\
	&
	&
	TWLSE &
	\begin{tabular}[c]{@{}c@{}}0.999\\ (0.956)\end{tabular} &
	\begin{tabular}[c]{@{}c@{}}1.000\\ (0.908)\end{tabular} &
	\begin{tabular}[c]{@{}c@{}}1.001\\ (0.922)\end{tabular} &
	\begin{tabular}[c]{@{}c@{}}1.000\\ (0.924)\end{tabular} &
	\begin{tabular}[c]{@{}c@{}}1.000\\ (0.932)\end{tabular} &
	\begin{tabular}[c]{@{}c@{}}1.001\\ (0.938)\end{tabular} &
	&
	&
	TWLSE &
	\begin{tabular}[c]{@{}c@{}}1.000\\ (0.952)\end{tabular} &
	\begin{tabular}[c]{@{}c@{}}1.000\\ (0.92)\end{tabular} &
	\begin{tabular}[c]{@{}c@{}}1.000\\ (0.928)\end{tabular} &
	\begin{tabular}[c]{@{}c@{}}1.000\\ (0.912)\end{tabular} &
	\begin{tabular}[c]{@{}c@{}}1.000\\ (0.912)\end{tabular} &
	\begin{tabular}[c]{@{}c@{}}1.000\\ (0.934)\end{tabular} \\ \cline{2-9} \cline{11-18}
	&
	\multirow{3}{*}{40} &
	OS &
	0.074 &
	0.291 &
	0.340 &
	0.319 &
	0.327 &
	0.316 &
	&
	\multirow{3}{*}{40} &
	OS &
	0.974 &
	0.976 &
	0.976 &
	1.001 &
	1.001 &
	0.992 \\
	&
	&
	WLSE &
	\begin{tabular}[c]{@{}c@{}}0.774\\ (0.888)\end{tabular} &
	\begin{tabular}[c]{@{}c@{}}0.986\\ (0.910)\end{tabular} &
	\begin{tabular}[c]{@{}c@{}}0.965\\ (0.920)\end{tabular} &
	\begin{tabular}[c]{@{}c@{}}0.962\\ (0.922)\end{tabular} &
	\begin{tabular}[c]{@{}c@{}}0.960\\ (0.934)\end{tabular} &
	\begin{tabular}[c]{@{}c@{}}0.964\\ (0.938)\end{tabular} &
	&
	&
	WLSE &
	\begin{tabular}[c]{@{}c@{}}0.907\\ (0.922)\end{tabular} &
	\begin{tabular}[c]{@{}c@{}}1.001\\ (0.916)\end{tabular} &
	\begin{tabular}[c]{@{}c@{}}1.000\\ (0.93)\end{tabular} &
	\begin{tabular}[c]{@{}c@{}}1.002\\ (0.91)\end{tabular} &
	\begin{tabular}[c]{@{}c@{}}0.996\\ (0.902)\end{tabular} &
	\begin{tabular}[c]{@{}c@{}}0.99\\ (0.936)\end{tabular} \\
	&
	&
	TWLSE &
	\begin{tabular}[c]{@{}c@{}}1.002\\ (0.958)\end{tabular} &
	\begin{tabular}[c]{@{}c@{}}1.001\\ (0.906)\end{tabular} &
	\begin{tabular}[c]{@{}c@{}}1.002\\ (0.920)\end{tabular} &
	\begin{tabular}[c]{@{}c@{}}0.999\\ (0.920)\end{tabular} &
	\begin{tabular}[c]{@{}c@{}}1.000\\ (0.932)\end{tabular} &
	\begin{tabular}[c]{@{}c@{}}1.000\\ (0.938)\end{tabular} &
	&
	&
	TWLSE &
	\begin{tabular}[c]{@{}c@{}}1.000\\ (0.952)\end{tabular} &
	\begin{tabular}[c]{@{}c@{}}1.000\\ (0.92)\end{tabular} &
	\begin{tabular}[c]{@{}c@{}}1.000\\ (0.928)\end{tabular} &
	\begin{tabular}[c]{@{}c@{}}1.000\\ (0.912)\end{tabular} &
	\begin{tabular}[c]{@{}c@{}}1.000\\ (0.912)\end{tabular} &
	\begin{tabular}[c]{@{}c@{}}1.000\\ (0.934)\end{tabular} \\ \hline
	\multirow{9}{*}{4} &
	\multirow{3}{*}{10} &
	OS &
	0.986 &
	0.992 &
	0.994 &
	0.994 &
	1.000 &
	0.998 &
	\multirow{9}{*}{20} &
	\multirow{3}{*}{10} &
	OS &
	0.99 &
	1.000 &
	0.997 &
	1.001 &
	0.999 &
	0.991 \\
	&
	&
	WLSE &
	\begin{tabular}[c]{@{}c@{}}1.006\\ (0.942)\end{tabular} &
	\begin{tabular}[c]{@{}c@{}}1.002\\ (0.946)\end{tabular} &
	\begin{tabular}[c]{@{}c@{}}1.002\\ (0.912)\end{tabular} &
	\begin{tabular}[c]{@{}c@{}}1.002\\ (0.922)\end{tabular} &
	\begin{tabular}[c]{@{}c@{}}1.002\\ (0.894)\end{tabular} &
	\begin{tabular}[c]{@{}c@{}}1.002\\ (0.914)\end{tabular} &
	&
	&
	WLSE &
	\begin{tabular}[c]{@{}c@{}}1.002\\ (0.964)\end{tabular} &
	\begin{tabular}[c]{@{}c@{}}1.000\\ (0.910)\end{tabular} &
	\begin{tabular}[c]{@{}c@{}}1.001\\ (0.922)\end{tabular} &
	\begin{tabular}[c]{@{}c@{}}1.000\\ (0.914)\end{tabular} &
	\begin{tabular}[c]{@{}c@{}}0.999\\ (0.948)\end{tabular} &
	\begin{tabular}[c]{@{}c@{}}1.001\\ (0.948)\end{tabular} \\
	&
	&
	TWLSE &
	\begin{tabular}[c]{@{}c@{}}1.000\\ (0.944)\end{tabular} &
	\begin{tabular}[c]{@{}c@{}}1.000\\ (0.942)\end{tabular} &
	\begin{tabular}[c]{@{}c@{}}1.000\\ (0.910)\end{tabular} &
	\begin{tabular}[c]{@{}c@{}}1.000\\ (0.922)\end{tabular} &
	\begin{tabular}[c]{@{}c@{}}1.000\\ (0.900)\end{tabular} &
	\begin{tabular}[c]{@{}c@{}}1.000\\ (0.920)\end{tabular} &
	&
	&
	TWLSE &
	\begin{tabular}[c]{@{}c@{}}1.000\\ (0.966)\end{tabular} &
	\begin{tabular}[c]{@{}c@{}}1.000\\ (0.910)\end{tabular} &
	\begin{tabular}[c]{@{}c@{}}1.000\\ (0.926)\end{tabular} &
	\begin{tabular}[c]{@{}c@{}}1.000\\ (0.912)\end{tabular} &
	\begin{tabular}[c]{@{}c@{}}1.000\\ (0.946)\end{tabular} &
	\begin{tabular}[c]{@{}c@{}}1.000\\ (0.942)\end{tabular} \\ \cline{2-9} \cline{11-18}
	&
	\multirow{3}{*}{20} &
	OS &
	0.959 &
	0.985 &
	0.975 &
	0.982 &
	0.982 &
	0.984 &
	&
	\multirow{3}{*}{20} &
	OS &
	0.982 &
	0.993 &
	0.995 &
	0.998 &
	0.992 &
	0.992 \\
	&
	&
	WLSE &
	\begin{tabular}[c]{@{}c@{}}0.984\\ (0.94)\end{tabular} &
	\begin{tabular}[c]{@{}c@{}}1.004\\ (0.946)\end{tabular} &
	\begin{tabular}[c]{@{}c@{}}1.004\\ (0.908)\end{tabular} &
	\begin{tabular}[c]{@{}c@{}}1.006\\ (0.924)\end{tabular} &
	\begin{tabular}[c]{@{}c@{}}1.002\\ (0.894)\end{tabular} &
	\begin{tabular}[c]{@{}c@{}}1.005\\ (0.912)\end{tabular} &
	&
	&
	WLSE &
	\begin{tabular}[c]{@{}c@{}}0.997\\ (0.966)\end{tabular} &
	\begin{tabular}[c]{@{}c@{}}1.001\\ (0.912)\end{tabular} &
	\begin{tabular}[c]{@{}c@{}}1.001\\ (0.920)\end{tabular} &
	\begin{tabular}[c]{@{}c@{}}1.000\\ (0.914)\end{tabular} &
	\begin{tabular}[c]{@{}c@{}}0.998\\ (0.946)\end{tabular} &
	\begin{tabular}[c]{@{}c@{}}1.001\\ (0.946)\end{tabular} \\
	&
	&
	TWLSE &
	\begin{tabular}[c]{@{}c@{}}1..000\\ (0.944)\end{tabular} &
	\begin{tabular}[c]{@{}c@{}}1.000\\ (0.942)\end{tabular} &
	\begin{tabular}[c]{@{}c@{}}1.000\\ (0.91)\end{tabular} &
	\begin{tabular}[c]{@{}c@{}}1.000\\ (0.922)\end{tabular} &
	\begin{tabular}[c]{@{}c@{}}1.000\\ (0.9)\end{tabular} &
	\begin{tabular}[c]{@{}c@{}}1.000\\ (0.92)\end{tabular} &
	&
	&
	TWLSE &
	\begin{tabular}[c]{@{}c@{}}1.000\\ (0.964)\end{tabular} &
	\begin{tabular}[c]{@{}c@{}}1.000\\ (0.91)\end{tabular} &
	\begin{tabular}[c]{@{}c@{}}1.000\\ (0.926)\end{tabular} &
	\begin{tabular}[c]{@{}c@{}}1.000\\ (0.912)\end{tabular} &
	\begin{tabular}[c]{@{}c@{}}1.000\\ (0.946)\end{tabular} &
	\begin{tabular}[c]{@{}c@{}}1.000\\ (0.942)\end{tabular} \\ \cline{2-9} \cline{11-18}
	&
	\multirow{3}{*}{40} &
	OS &
	0.281 &
	0.920 &
	0.911 &
	0.917 &
	0.857 &
	0.741 &
	&
	\multirow{3}{*}{40} &
	OS &
	0.968 &
	0.991 &
	0.993 &
	0.995 &
	0.989 &
	0.987 \\
	&
	&
	WLSE &
	\begin{tabular}[c]{@{}c@{}}0.869\\ (0.902)\end{tabular} &
	\begin{tabular}[c]{@{}c@{}}1.001\\ (0.946)\end{tabular} &
	\begin{tabular}[c]{@{}c@{}}1.003\\ (0.908)\end{tabular} &
	\begin{tabular}[c]{@{}c@{}}1.001\\ (0.918)\end{tabular} &
	\begin{tabular}[c]{@{}c@{}}0.994\\ (0.9)\end{tabular} &
	\begin{tabular}[c]{@{}c@{}}0.999\\ (0.906)\end{tabular} &
	&
	&
	WLSE &
	\begin{tabular}[c]{@{}c@{}}0.962\\ (0.96)\end{tabular} &
	\begin{tabular}[c]{@{}c@{}}1.001\\ (0.916)\end{tabular} &
	\begin{tabular}[c]{@{}c@{}}1.002\\ (0.918)\end{tabular} &
	\begin{tabular}[c]{@{}c@{}}0.998\\ (0.916)\end{tabular} &
	\begin{tabular}[c]{@{}c@{}}0.994\\ (0.942)\end{tabular} &
	\begin{tabular}[c]{@{}c@{}}1.000\\ (0.946)\end{tabular} \\
	&
	&
	TWLSE &
	\begin{tabular}[c]{@{}c@{}}1.001\\ (0.94)\end{tabular} &
	\begin{tabular}[c]{@{}c@{}}1.000\\ (0.942)\end{tabular} &
	\begin{tabular}[c]{@{}c@{}}1.000\\ (0.908)\end{tabular} &
	\begin{tabular}[c]{@{}c@{}}1.000\\ (0.922)\end{tabular} &
	\begin{tabular}[c]{@{}c@{}}1.000\\ (0.898)\end{tabular} &
	\begin{tabular}[c]{@{}c@{}}1.000\\ (0.920)\end{tabular} &
	&
	&
	TWLSE &
	\begin{tabular}[c]{@{}c@{}}1.000\\ (0.964)\end{tabular} &
	\begin{tabular}[c]{@{}c@{}}1.000\\ (0.910)\end{tabular} &
	\begin{tabular}[c]{@{}c@{}}1.000\\ (0.926)\end{tabular} &
	\begin{tabular}[c]{@{}c@{}}1.000\\ (0.912)\end{tabular} &
	\begin{tabular}[c]{@{}c@{}}1.000\\ (0.946)\end{tabular} &
	\begin{tabular}[c]{@{}c@{}}1.000\\ (0.942)\end{tabular} \\ \hline
\end{tabular}
}
\end{sidewaystable}

\begin{sidewaystable}[thp]\footnotesize
		\centering
		\caption{``REE" results for Example 2 with 500 replications. The performances are evaluated for different sample sizes $N(\times10^3)$ and numbers of workers $K$. The corresponding CPs are displayed in parentheses.}\label{table:example2}
	\scalebox{0.9}{
\begin{tabular}{c|c|ccccccc|c|c|ccccccc}
	\hline
	$N$ &
	$K$ &
	Estimation &
	$\rho$ &
	$\beta_1$ &
	$\beta_2$ &
	$\beta_3$ &
	$\beta_4$ &
	$\beta_5$ &
	$N$ &
	$K$ &
	Estimation &
	$\rho$ &
	$\beta_1$ &
	$\beta_2$ &
	$\beta_3$ &
	$\beta_4$ &
	$\beta_5$ \\ \hline
	\multirow{9}{*}{2} &
	\multirow{3}{*}{10} &
	OS &
	0.520 &
	0.904 &
	0.923 &
	0.992 &
	0.964 &
	0.944 &
	\multirow{9}{*}{10} &
	\multirow{3}{*}{10} &
	OS &
	0.990 &
	0.991 &
	0.999 &
	0.998 &
	0.996 &
	0.997 \\
	&
	&
	WLSE &
	\begin{tabular}[c]{@{}c@{}}0.933\\ (0.926)\end{tabular} &
	\begin{tabular}[c]{@{}c@{}}1.001\\ (0.910)\end{tabular} &
	\begin{tabular}[c]{@{}c@{}}1.001\\ (0.918)\end{tabular} &
	\begin{tabular}[c]{@{}c@{}}1.005\\ (0.928)\end{tabular} &
	\begin{tabular}[c]{@{}c@{}}0.998\\ (0.952)\end{tabular} &
	\begin{tabular}[c]{@{}c@{}}0.999\\ (0.934)\end{tabular} &
	&
	&
	WLSE &
	\begin{tabular}[c]{@{}c@{}}1.000\\ (0.944)\end{tabular} &
	\begin{tabular}[c]{@{}c@{}}1.001\\ (0.940)\end{tabular} &
	\begin{tabular}[c]{@{}c@{}}1.001\\ (0.918)\end{tabular} &
	\begin{tabular}[c]{@{}c@{}}0.999\\ (0.924)\end{tabular} &
	\begin{tabular}[c]{@{}c@{}}1.002\\ (0.954)\end{tabular} &
	\begin{tabular}[c]{@{}c@{}}1.002\\ (0.942)\end{tabular} \\
	&
	&
	TWLSE &
	\begin{tabular}[c]{@{}c@{}}0.998\\ (0.940)\end{tabular} &
	\begin{tabular}[c]{@{}c@{}}1.000\\ (0.906)\end{tabular} &
	\begin{tabular}[c]{@{}c@{}}1.000\\ (0.918)\end{tabular} &
	\begin{tabular}[c]{@{}c@{}}0.999\\ (0.928)\end{tabular} &
	\begin{tabular}[c]{@{}c@{}}1.000\\ (0.954)\end{tabular} &
	\begin{tabular}[c]{@{}c@{}}1.000\\ (0.932)\end{tabular} &
	&
	&
	TWLSE &
	\begin{tabular}[c]{@{}c@{}}1.000\\ (0.944)\end{tabular} &
	\begin{tabular}[c]{@{}c@{}}1.000\\ (0.938)\end{tabular} &
	\begin{tabular}[c]{@{}c@{}}1.000\\ (0.918)\end{tabular} &
	\begin{tabular}[c]{@{}c@{}}1.000\\ (0.922)\end{tabular} &
	\begin{tabular}[c]{@{}c@{}}1.000\\ (0.950)\end{tabular} &
	\begin{tabular}[c]{@{}c@{}}1.000\\ (0.946)\end{tabular} \\ \cline{2-9} \cline{11-18}
	&
	\multirow{3}{*}{20} &
	OS &
	0.312 &
	0.710 &
	0.783 &
	0.752 &
	0.679 &
	0.725 &
	&
	\multirow{3}{*}{20} &
	OS &
	0.982 &
	0.990 &
	0.997 &
	0.997 &
	0.995 &
	0.985 \\
	&
	&
	WLSE &
	\begin{tabular}[c]{@{}c@{}}0.785\\ (0.880)\end{tabular} &
	\begin{tabular}[c]{@{}c@{}}0.988\\ (0.902)\end{tabular} &
	\begin{tabular}[c]{@{}c@{}}0.979\\ (0.916)\end{tabular} &
	\begin{tabular}[c]{@{}c@{}}0.998\\ (0.926)\end{tabular} &
	\begin{tabular}[c]{@{}c@{}}0.984\\ (0.948)\end{tabular} &
	\begin{tabular}[c]{@{}c@{}}0.963\\ (0.922)\end{tabular} &
	&
	&
	WLSE &
	\begin{tabular}[c]{@{}c@{}}0.980\\ (0.930)\end{tabular} &
	\begin{tabular}[c]{@{}c@{}}1.002\\ (0.938)\end{tabular} &
	\begin{tabular}[c]{@{}c@{}}1.002\\ (0.916)\end{tabular} &
	\begin{tabular}[c]{@{}c@{}}0.998\\ (0.924)\end{tabular} &
	\begin{tabular}[c]{@{}c@{}}1.004\\ (0.954)\end{tabular} &
	\begin{tabular}[c]{@{}c@{}}1.003\\ (0.940)\end{tabular} \\
	&
	&
	TWLSE &
	\begin{tabular}[c]{@{}c@{}}0.995\\ (0.938)\end{tabular} &
	\begin{tabular}[c]{@{}c@{}}1.000\\ (0.906)\end{tabular} &
	\begin{tabular}[c]{@{}c@{}}0.999\\ (0.916)\end{tabular} &
	\begin{tabular}[c]{@{}c@{}}1.000\\ (0.928)\end{tabular} &
	\begin{tabular}[c]{@{}c@{}}0.999\\ (0.954)\end{tabular} &
	\begin{tabular}[c]{@{}c@{}}0.998\\ (0.934)\end{tabular} &
	&
	&
	TWLSE &
	\begin{tabular}[c]{@{}c@{}}1.000\\ (0.944)\end{tabular} &
	\begin{tabular}[c]{@{}c@{}}1.000\\ (0.938)\end{tabular} &
	\begin{tabular}[c]{@{}c@{}}1.000\\ (0.918)\end{tabular} &
	\begin{tabular}[c]{@{}c@{}}1.000\\ (0.922)\end{tabular} &
	\begin{tabular}[c]{@{}c@{}}1.000\\ (0.95)\end{tabular} &
	\begin{tabular}[c]{@{}c@{}}1.000\\ (0.944)\end{tabular} \\ \cline{2-9} \cline{11-18}
	&
	\multirow{3}{*}{40} &
	OS &
	0.070 &
	0.246 &
	0.241 &
	0.197 &
	0.177 &
	0.242 &
	&
	\multirow{3}{*}{40} &
	OS &
	0.952 &
	0.978 &
	0.989 &
	0.989 &
	0.992 &
	0.975 \\
	&
	&
	WLSE &
	\begin{tabular}[c]{@{}c@{}}0.320\\ (0.652)\end{tabular} &
	\begin{tabular}[c]{@{}c@{}}0.815\\ (0.884)\end{tabular} &
	\begin{tabular}[c]{@{}c@{}}0.782\\ (0.864)\end{tabular} &
	\begin{tabular}[c]{@{}c@{}}0.631\\ (0.898)\end{tabular} &
	\begin{tabular}[c]{@{}c@{}}0.812\\ (0.914)\end{tabular} &
	\begin{tabular}[c]{@{}c@{}}0.785\\ (0.886)\end{tabular} &
	&
	&
	WLSE &
	\begin{tabular}[c]{@{}c@{}}0.893\\ (0.906)\end{tabular} &
	\begin{tabular}[c]{@{}c@{}}1.004\\ (0.940)\end{tabular} &
	\begin{tabular}[c]{@{}c@{}}1.003\\ (0.918)\end{tabular} &
	\begin{tabular}[c]{@{}c@{}}0.996\\ (0.924)\end{tabular} &
	\begin{tabular}[c]{@{}c@{}}1.007\\ (0.952)\end{tabular} &
	\begin{tabular}[c]{@{}c@{}}1.002\\ (0.936)\end{tabular} \\
	&
	&
	TWLSE &
	\begin{tabular}[c]{@{}c@{}}0.859\\ (0.906)\end{tabular} &
	\begin{tabular}[c]{@{}c@{}}1.002\\ (0.902)\end{tabular} &
	\begin{tabular}[c]{@{}c@{}}0.980\\ (0.914)\end{tabular} &
	\begin{tabular}[c]{@{}c@{}}0.962\\ (0.916)\end{tabular} &
	\begin{tabular}[c]{@{}c@{}}0.933\\ (0.948)\end{tabular} &
	\begin{tabular}[c]{@{}c@{}}0.919\\ (0.926)\end{tabular} &
	&
	&
	TWLSE &
	\begin{tabular}[c]{@{}c@{}}1.000\\ (0.942)\end{tabular} &
	\begin{tabular}[c]{@{}c@{}}1.000\\ (0.938)\end{tabular} &
	\begin{tabular}[c]{@{}c@{}}1.000\\ (0.918)\end{tabular} &
	\begin{tabular}[c]{@{}c@{}}1.000\\ (0.922)\end{tabular} &
	\begin{tabular}[c]{@{}c@{}}1.000\\ (0.95)\end{tabular} &
	\begin{tabular}[c]{@{}c@{}}1.000\\ (0.942)\end{tabular} \\ \hline
	\multirow{9}{*}{4} &
	\multirow{3}{*}{10} &
	OS &
	0.981 &
	0.995 &
	0.990 &
	0.986 &
	0.988 &
	1.001 &
	\multirow{9}{*}{20} &
	\multirow{3}{*}{10} &
	OS &
	0.997 &
	0.998 &
	0.998 &
	0.994 &
	0.999 &
	1.001 \\
	&
	&
	WLSE &
	\begin{tabular}[c]{@{}c@{}}0.997\\ (0.950)\end{tabular} &
	\begin{tabular}[c]{@{}c@{}}1.001\\ (0.910)\end{tabular} &
	\begin{tabular}[c]{@{}c@{}}1.004\\ (0.910)\end{tabular} &
	\begin{tabular}[c]{@{}c@{}}1.003\\ (0.926)\end{tabular} &
	\begin{tabular}[c]{@{}c@{}}1.002\\ (0.926)\end{tabular} &
	\begin{tabular}[c]{@{}c@{}}1.001\\ (0.950)\end{tabular} &
	&
	&
	WLSE &
	\begin{tabular}[c]{@{}c@{}}1.003\\ (0.932)\end{tabular} &
	\begin{tabular}[c]{@{}c@{}}1.001\\ (0.952)\end{tabular} &
	\begin{tabular}[c]{@{}c@{}}1.000\\ (0.902)\end{tabular} &
	\begin{tabular}[c]{@{}c@{}}1.001\\ (0.936)\end{tabular} &
	\begin{tabular}[c]{@{}c@{}}0.999\\ (0.944)\end{tabular} &
	\begin{tabular}[c]{@{}c@{}}1.000\\ (0.932)\end{tabular} \\
	&
	&
	TWLSE &
	\begin{tabular}[c]{@{}c@{}}1.000\\ (0.946)\end{tabular} &
	\begin{tabular}[c]{@{}c@{}}1.000\\ (0.910)\end{tabular} &
	\begin{tabular}[c]{@{}c@{}}1.000\\ (0.908)\end{tabular} &
	\begin{tabular}[c]{@{}c@{}}1.000\\ (0.928)\end{tabular} &
	\begin{tabular}[c]{@{}c@{}}1.000\\ (0.920)\end{tabular} &
	\begin{tabular}[c]{@{}c@{}}1.000\\ (0.942)\end{tabular} &
	&
	&
	TWLSE &
	\begin{tabular}[c]{@{}c@{}}1.000\\ (0.936)\end{tabular} &
	\begin{tabular}[c]{@{}c@{}}1.000\\ (0.952)\end{tabular} &
	\begin{tabular}[c]{@{}c@{}}1.000\\ (0.904)\end{tabular} &
	\begin{tabular}[c]{@{}c@{}}1.000\\ (0.938)\end{tabular} &
	\begin{tabular}[c]{@{}c@{}}1.000\\ (0.944)\end{tabular} &
	\begin{tabular}[c]{@{}c@{}}1.000\\ (0.934)\end{tabular} \\ \cline{2-9} \cline{11-18}
	&
	\multirow{3}{*}{20} &
	OS &
	0.928 &
	0.986 &
	0.977 &
	0.971 &
	0.968 &
	0.981 &
	&
	\multirow{3}{*}{20} &
	OS &
	0.993 &
	0.993 &
	0.996 &
	0.993 &
	1.000 &
	0.998 \\
	&
	&
	WLSE &
	\begin{tabular}[c]{@{}c@{}}0.937\\ (0.944)\end{tabular} &
	\begin{tabular}[c]{@{}c@{}}1.001\\ (0.912)\end{tabular} &
	\begin{tabular}[c]{@{}c@{}}1.006\\ (0.910)\end{tabular} &
	\begin{tabular}[c]{@{}c@{}}1.003\\ (0.926)\end{tabular} &
	\begin{tabular}[c]{@{}c@{}}1.004\\ (0.930)\end{tabular} &
	\begin{tabular}[c]{@{}c@{}}0.997\\ (0.948)\end{tabular} &
	&
	&
	WLSE &
	\begin{tabular}[c]{@{}c@{}}0.997\\ (0.926)\end{tabular} &
	\begin{tabular}[c]{@{}c@{}}1.001\\ (0.952)\end{tabular} &
	\begin{tabular}[c]{@{}c@{}}1.001\\ (0.900)\end{tabular} &
	\begin{tabular}[c]{@{}c@{}}1.001\\ (0.932)\end{tabular} &
	\begin{tabular}[c]{@{}c@{}}0.998\\ (0.942)\end{tabular} &
	\begin{tabular}[c]{@{}c@{}}0.999\\ (0.932)\end{tabular} \\
	&
	&
	TWLSE &
	\begin{tabular}[c]{@{}c@{}}1.000\\ (0.946)\end{tabular} &
	\begin{tabular}[c]{@{}c@{}}1.000\\ (0.912)\end{tabular} &
	\begin{tabular}[c]{@{}c@{}}1.000\\ (0.906)\end{tabular} &
	\begin{tabular}[c]{@{}c@{}}1.000\\ (0.928)\end{tabular} &
	\begin{tabular}[c]{@{}c@{}}1.000\\ (0.918)\end{tabular} &
	\begin{tabular}[c]{@{}c@{}}1.000\\ (0.942)\end{tabular} &
	&
	&
	TWLSE &
	\begin{tabular}[c]{@{}c@{}}1.000\\ (0.936)\end{tabular} &
	\begin{tabular}[c]{@{}c@{}}1.000\\ (0.952)\end{tabular} &
	\begin{tabular}[c]{@{}c@{}}1.000\\ (0.904)\end{tabular} &
	\begin{tabular}[c]{@{}c@{}}1.000\\ (0.938)\end{tabular} &
	\begin{tabular}[c]{@{}c@{}}1.000\\ (0.944)\end{tabular} &
	\begin{tabular}[c]{@{}c@{}}1.000\\ (0.934)\end{tabular} \\ \cline{2-9} \cline{11-18}
	&
	\multirow{3}{*}{40} &
	OS &
	0.252 &
	0.739 &
	0.722 &
	0.680 &
	0.684 &
	0.685 &
	&
	\multirow{3}{*}{40} &
	OS &
	0.981 &
	0.994 &
	0.992 &
	0.982 &
	0.998 &
	0.988 \\
	&
	&
	WLSE &
	\begin{tabular}[c]{@{}c@{}}0.673\\ (0.852)\end{tabular} &
	\begin{tabular}[c]{@{}c@{}}0.983\\ (0.908)\end{tabular} &
	\begin{tabular}[c]{@{}c@{}}0.977\\ (0.914)\end{tabular} &
	\begin{tabular}[c]{@{}c@{}}0.959\\ (0.912)\end{tabular} &
	\begin{tabular}[c]{@{}c@{}}0.976\\ (0.916)\end{tabular} &
	\begin{tabular}[c]{@{}c@{}}0.956\\ (0.938)\end{tabular} &
	&
	&
	WLSE &
	\begin{tabular}[c]{@{}c@{}}0.954\\ (0.926)\end{tabular} &
	\begin{tabular}[c]{@{}c@{}}1.003\\ (0.952)\end{tabular} &
	\begin{tabular}[c]{@{}c@{}}1.002\\ (0.904)\end{tabular} &
	\begin{tabular}[c]{@{}c@{}}1.002\\ (0.928)\end{tabular} &
	\begin{tabular}[c]{@{}c@{}}0.996\\ (0.934)\end{tabular} &
	\begin{tabular}[c]{@{}c@{}}0.996\\ (0.930)\end{tabular} \\
	&
	&
	TWLSE &
	\begin{tabular}[c]{@{}c@{}}1.001\\ (0.944)\end{tabular} &
	\begin{tabular}[c]{@{}c@{}}1.000\\ (0.912)\end{tabular} &
	\begin{tabular}[c]{@{}c@{}}1.000\\ (0.908)\end{tabular} &
	\begin{tabular}[c]{@{}c@{}}1.001\\ (0.926)\end{tabular} &
	\begin{tabular}[c]{@{}c@{}}1.000\\ (0.920)\end{tabular} &
	\begin{tabular}[c]{@{}c@{}}1.000\\ (0.942)\end{tabular} &
	&
	&
	TWLSE &
	\begin{tabular}[c]{@{}c@{}}1.000\\ (0.936)\end{tabular} &
	\begin{tabular}[c]{@{}c@{}}1.000\\ (0.952)\end{tabular} &
	\begin{tabular}[c]{@{}c@{}}1.000\\ (0.904)\end{tabular} &
	\begin{tabular}[c]{@{}c@{}}1.000\\ (0.936)\end{tabular} &
	\begin{tabular}[c]{@{}c@{}}1.000\\ (0.944)\end{tabular} &
	\begin{tabular}[c]{@{}c@{}}1.000\\ (0.934)\end{tabular} \\ \hline
\end{tabular}
}
\end{sidewaystable}

\section{A YELP DATA ANALYSIS}\label{sec:realdata}

In this section, we apply the proposed method to a Yelp dataset collected from Yelp's official public website (\url{https://www.yelp.com/dataset/}).
As one of the most popular online guides for evaluating and recommending a large range of businesses, Yelp has accumulated millions of users by 2020.
The objective here is to investigate how the Yelp user's friends' behaviors influence the user's review.
In the following, we first present a description of the data, and then implement the proposed method on the dataset to illustrate the usefulness of our distributed estimation and inference procedure.

\subsection{Data Description}

\begin{figure}[htpb!]
	\begin{center}
		\includegraphics[width=0.7\textwidth]{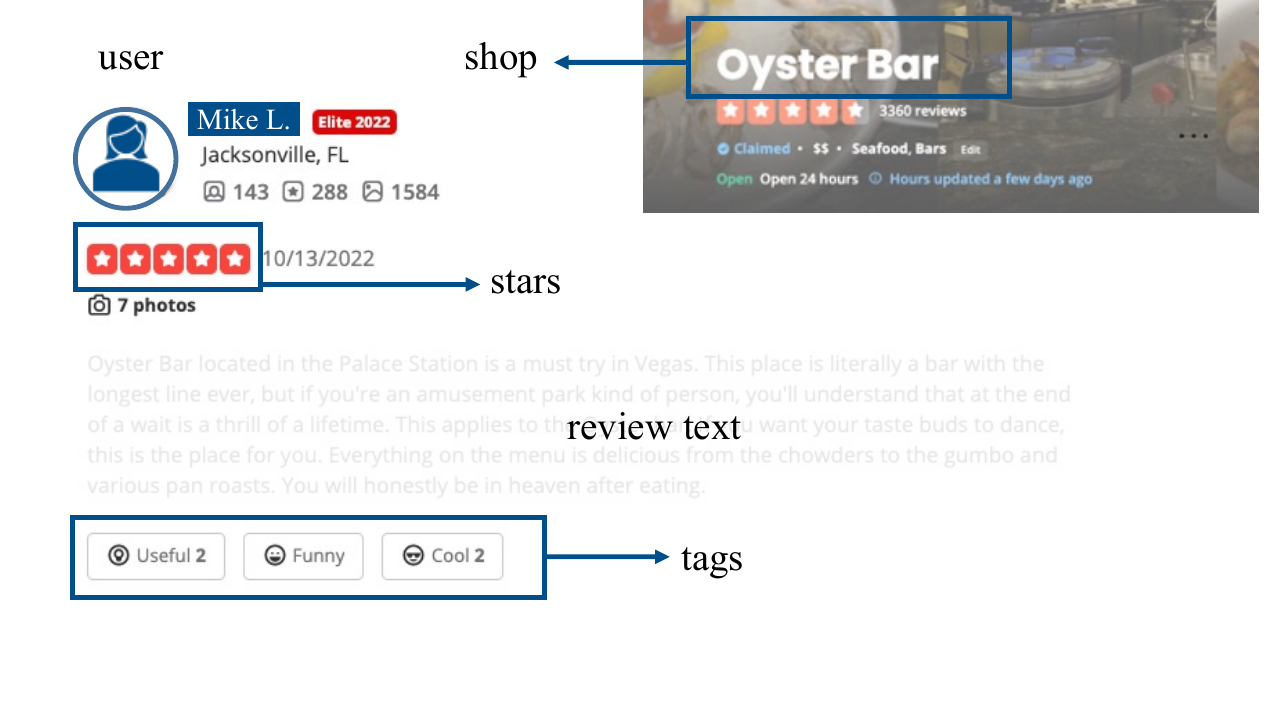}
		\caption{\small One review from a user for a shop named ``Oyster Bar". It contains the user and shop information, shop rating, the review text, and the number of three tags (i.e., ``useful", ``funny", and ``cool") for this review.}
		\label{fig:yelp_page}
	\end{center}
\end{figure}

The Yelp dataset is collected from 12th October, 2004 to 12th February, 2020 and consists of
$N = 945,140$ users.
For the modeling purpose, we record the network relationship, users' characteristic data, and user-shop reviewing data.
The reviewing data includes the tags assigned to each review, namely, ``useful", ``funny", and ``cool".
See Figure \ref{fig:yelp_page} for an illustration of one user review.
As shown in the figure, the user rated the restaurant ``Oyster Bar" with five stars.
In addition, this user's review comment received two ``useful" tags and two ``cool" tags from other users.

\begin{figure}[htp]
	\setlength{\lineskip}{0pt}
	\centering
		\includegraphics[width=0.8\textwidth]{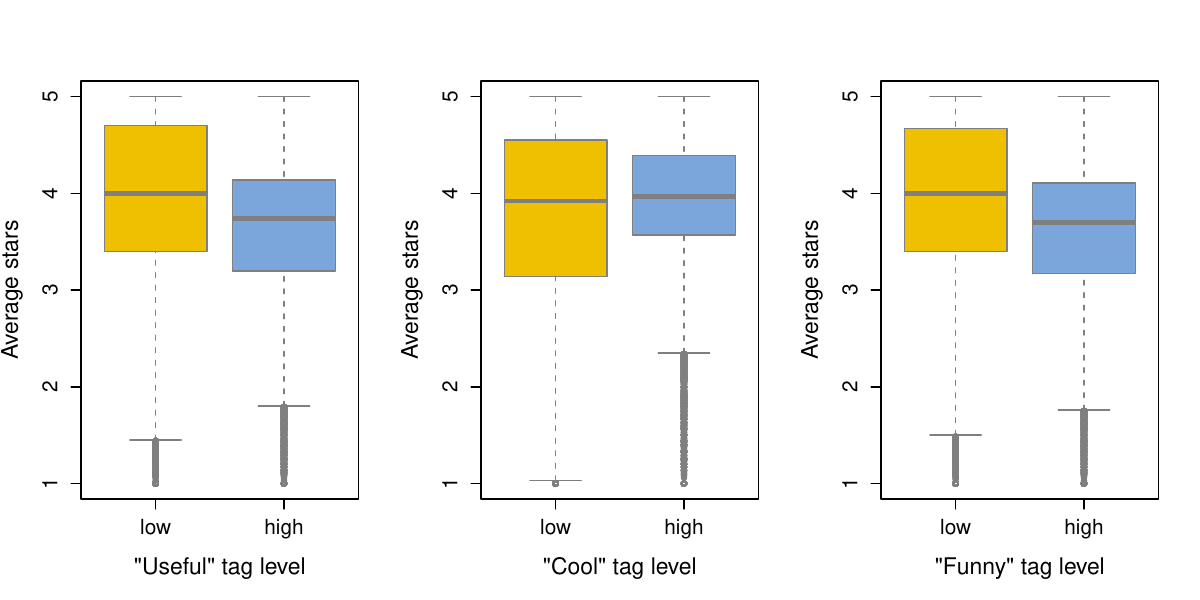}
		\caption{\small Boxplots of response variable $Y_{i}$ with regard to the three tag covariates, which are ``useful'', ``cool'' and ``funny''. The covariates are split into ``high" and ``low" according to their mean values.}
		\label{fig:yelp_des}
\end{figure}

To construct the adjacency matrix $\bA$, we set $a_{ij} = 1$ if user $j$ is a friend of user $i$ on Yelp.
This leads to a network with 945,140 nodes and more than 19 million edges. The network density is expressed as $\sum_{i,j}a_{ij}/\{N(N-1)\} = 4.26 \times 10^{-5}$, which is extremely sparse.
The response variable $Y_{i}$ is defined as the averaged ``stars" scores given by user $i$, which reflects the average review quality delivered by this user.
Then, we consider four meaningful covariates for each user.
First, we use $X_{i,\text{use}}$ (useful), $X_{i, \text{cool}}$ (cool), and $X_{i, \text{fun}}$ (funny) to describe the popularity of the users' reviews.
Using the tag ``useful" as an example,
if the user's comment was found to be useful by another user $j$, then user $j$ will tag {\it useful} on the comment from user $i$.
The cumulative number of ``useful" tags on each comment reflects
how much the comment is appreciated by other users.
Then, we calculate $X_{i,\text{use}}$ as the average ``useful" tags for each user, i.e., the total number of ``useful" tags divided by the number of reviews of this user.
The covariates $X_{i, \text{cool}}$ and $X_{i, \text{fun}}$ are calculated in the same way using ``cool" and ``funny" tags, respectively.
Additionally, we include $X_{i, \text{fol}}$ as the total number of followers for each user since it could reflect the social activeness on the Yelp platform.
We visualize the relationship between the response and tag-related covariates in Figure \ref{fig:yelp_des}, where the covariates are split by the mean value.
According to Figure \ref{fig:yelp_des}, we find that users with more ``useful'' and
``funny" tags tend to rate lower scores than others.
Then, all of the variables are standardized with a zero mean and a unit variance for later modeling.
Subsequently, we use the model \eqref{sar}, and implement Algorithm \ref{alg:dsar} to estimate the network effect from friends' average review scores ($\rho$), and the covariates effect ($\bbeta$) on the user's review average scores.

{
\subsection{Spark System Implementation and Results}\label{subsec:yelp_res}
To evaluate the numerical performance of our proposed WLSE and TWLSE, we establish a Spark-on-YARN cluster, which is a commonly used deployment for the Spark system.
Our cluster contains a master node (i.e., driver) and two worker nodes. The master node has 32 virtual cores with 256 GB of RAM, and each worker node has 8 virtual cores with 64 GB of RAM.
Collectively, this configuration provides a total of 48 virtual cores and 384 GB of RAM in the entire distributed system.
Thus, we apply 12 executors from the scheduler, and each executor has four virtual cores with 20 GB of RAM.
On the system, our dataset is randomly split into 48 subgroups with each subgroup containing approximately $19,690$ individuals.
Then, for each partition, we run the local estimation algorithm on a fixed executor and finally aggregate the results from all executors to obtain the final result.
We further remark that the proposed method is not restricted to any
particular choice of hardware or software, and we provide an implementation with GPU based system in \ref{subsec:gpu_implement}.
To speed up the algorithm, we utilize the sparse random projection matrices for statistical inference \citep{johnson1984extensions, achlioptas2001database, li2006very}.

For the estimation results, we compare our algorithms with the non-distributed SAR global estimation method \citep{huang2019least} to show the differences and similarities among the three methods.
The non-distributed SAR global estimation method is executed on the master node, harnessing the complete computational potential of 32 cores and a memory capacity of 256 GB.
The estimation results are shown in Table \ref{tbl:realdata_results}.
One could see that the results of WLSE and TWLSE are similar, and much closer to the global estimate than the OS estimator.
Take the results of TWLSE for example,
the network effect $\wh\rho = 0.1120$ is significantly positive, which means that users' friends have a positive influence on users' review scores.
For the covariates, if the user's comments are more tagged as ``cool'', then this user tends to give an average higher rating toward the shops.
However, if the user's comments are more tagged as ``useful'' and ``funny'', the user is more likely to give a lower ``stars'' rating.
This is understandable because users are more likely to avoid making incorrect choices with the help of others' reviews. This phenomenon could explain why ``useful'' tag owners tend to give negative comments.
Hence, for the shops themselves, more attention should be paid on the reviews by customers who have more ``useful'' and ``funny'' tags than others.
Next, for a specific user who usually gives more comments on Yelp, he or she may give higher ratings. This indicates that users with more comments probably feel more satisfaction with the shops on Yelp.
Moreover, an interesting fact shows that users with more followers tend to give negative reviews. The shops may also need to pay more attention to these users since they could have higher network influences.
Lastly, we also conduct some model checking procedures for the residuals. We further apply the robust GMM for estimation by eliminating several possible endogenous variables. The details are provided in \ref{subsec:model_check}
in the supplementary material.
We would like to leave the study of the endogeneity issue as an important future topic.

Afterwards, to compare the computational efficiency, we calculate the time costs of the above methods, which are displayed in Table \ref{tbl:computation_time}.
To provide a clear illustration, we show the time cost of the computation and the worker-master communication.
The communication time is approximated by deducting the computational time on the master machine and the median computational time employed across the workers from the total runtime.
In addition, we report the computational cost for parameter estimation and statistical inference.
For parameter estimation, both the WLSE and TWLSE are much faster than the global estimator.
Specifically, the WLSE requires approximately 3.52 seconds for estimation in total, while the global estimation takes 109.20 seconds, which is around 30 times that of the WLSE.
Next, in terms of statistical inference, we set the projection dimension $d = [\log N]+1$ for both WLSE and TWLSE as in the simulation study.
We find that it takes around 41.9 seconds for both methods to complete the statistical inference.
Here, statistical inference consumes more computational time than estimation since it involves more complicated calculations.
However, for the global method, the direct inference procedure is infeasible with large-scale networks due to the memory constraints and the requirement for large storage spaces.
Even though it is implemented using the hard drive, it still needs to calculate matrices as $\bW^\top\bW$ and it consumes more than 6 TB on the hard drive and needs a large number of I/O procedures.
To partially address this concern, we take the advice of an anonymous
referee to conduct the global inference by splitting the matrices $\bW$ and $\bW^\top\bW$ into small chunks and load them sequentially into the RAM for computation.
We explain the implementation details in \ref{sec:chunk_infer}.
It takes more than 67 hours using the same machine described in Section \ref{subsec:comp}.
The $p$-values are also reported in Table \ref{tbl:realdata_results}, which is consistent with the inference result of TWLSE method.
As a consequence, the proposed distributed estimation and statistical inference framework is a more feasible choice when only limited computational resources are available.

\begin{table}[htbp]
	\centering
		\caption{\small Estimation and inference results of TWLSE, WLSE, OS, and the global method. The $p$-values of the estimated parameters are shown in parentheses.
The $p$-values of the global method is calculated using the chunk-based global inference procedure in \ref{sec:chunk_infer}.}
\scalebox{0.9}{
\begin{tabular}{c|cccccc}
	\hline
	\multirow{2}{*}{Method} & \multicolumn{6}{c}{Estimation}                                                                                                                                                                                                                                                                                                                                                                                                 \\ \cline{2-7}
	& $\wh\rho$                                                             & $\wh\beta_{\text{com}}$                                               & $\wh\beta_{\text{use}}$                                                & $\wh\beta_{\text{cool}}$                                              & $\wh\beta_{\text{fun}}$                                                & $\wh\beta_{\text{fol}}$                                                \\ \hline
	TWLSE                   & \begin{tabular}[c]{@{}c@{}}0.112\\ (\textless{}0.001)\end{tabular} & \begin{tabular}[c]{@{}c@{}}0.019\\ (\textless{}0.001)\end{tabular} & \begin{tabular}[c]{@{}c@{}}-0.426\\ (\textless{}0.001)\end{tabular} & \begin{tabular}[c]{@{}c@{}}0.480\\ (\textless{}0.001)\end{tabular}  & \begin{tabular}[c]{@{}c@{}}-0.064\\ (\textless{}0.001)\end{tabular} & \begin{tabular}[c]{@{}c@{}}-0.070\\ (\textless{}0.001)\end{tabular}  \\
	WLSE                    & \begin{tabular}[c]{@{}c@{}}0.098\\ (\textless{}0.001)\end{tabular} & \begin{tabular}[c]{@{}c@{}}0.018\\ (\textless{}0.001)\end{tabular} & \begin{tabular}[c]{@{}c@{}}-0.423\\ (\textless{}0.001)\end{tabular} & \begin{tabular}[c]{@{}c@{}}0.476\\ (\textless{}0.001)\end{tabular} & \begin{tabular}[c]{@{}c@{}}-0.064\\ (\textless{}0.001)\end{tabular} & \begin{tabular}[c]{@{}c@{}}-0.068\\ (\textless{}0.001)\end{tabular} \\\hline
	OS                      & 0.098                                                              & 0.023                                                              & -0.645                                                              & 0.611                                                              & -0.323                                                              & -0.038                                                              \\\hline
	Global                  & \begin{tabular}[c]{@{}c@{}}0.112\\ ({\textless{}0.001})\end{tabular} & \begin{tabular}[c]{@{}c@{}}0.019\\ ({\textless{}0.001})\end{tabular} & \begin{tabular}[c]{@{}c@{}}-0.426\\ ({\textless{}0.001})\end{tabular} & \begin{tabular}[c]{@{}c@{}}0.480\\ ({\textless{}0.001})\end{tabular}  & \begin{tabular}[c]{@{}c@{}}-0.064\\ ({\textless{}0.001})\end{tabular} & \begin{tabular}[c]{@{}c@{}}-0.071\\ ({\textless{}0.001})\end{tabular}  \\
\hline
\end{tabular}}
	\label{tbl:realdata_results}
\end{table}

\begin{table}[htbp]
	\centering
		\caption{\small Computation procedure and communication time cost (in seconds) of TWLSE, WLSE, OS, and the global method.
 The computational time of the global inference is evaluated using the chunk-based global inference procedure in \ref{sec:chunk_infer}.}
	\scalebox{0.9}{
	\begin{tabular}{c|cc|cc}
		\hline
		\multirow{2}{*}{Method} & \multicolumn{2}{c|}{Computation} & \multicolumn{2}{c}{Communication} \\ \cline{2-5}
		& Estimation      & Inference      & Estimation       & Inference      \\ \hline
		TWLSE                   & 3.5263          & 41.9173        & 21.6778          & 107.7718       \\
		WLSE                    & 3.5194          & 41.9069        & 12.1057          & 106.3102       \\
		OS                      & 3.5188          & -              & 12.1057          & -              \\
		Global                  & 109.1959        & {$2.42\times10^5$}              & -                & -              \\ \hline
	\end{tabular}}
	\label{tbl:computation_time}
\end{table}

} 

\section{CONCLUSION}\label{sec:conclusion}

In this paper, we propose a distributed estimation framework for the SAR model based on a least squares objective function.
Specifically, a distributed least squares approximation (DNLSA) method is developed. Then, we obtain a weighted least squares estimator (WLSE) using one-round communication between the master node and worker nodes in this system.
A refinement for a two-step estimator, namely, TWLSE, is further designed to reduce the estimation bias.
To make a valid statistical inference, we employ a random projection method to reduce the communication cost.
The asymptotic properties are derived for the two estimators.
In addition, the estimated asymptotic covariance is shown to be consistent
when the projection dimension is chosen appropriately.
This guarantees a valid statistical inference procedure with a low communication cost.
We illustrate the desirable performance of our proposed methods through several simulation studies and a real data example on the Yelp dataset.

Beyond the scope of our work, there are still some intriguing directions for future research.
First, the distributed estimation is designed for the SAR model based on the recent least squares estimation method  \citep{huang2019least,zhu2020multivariate}.
Accordingly, the distributed framework based on other popular estimation methods such as GMM and IV-based methods \citep{lin2010gmm, liu2017gmm, kelejian2004estimation, baltagi2015ec3sls, cohen2018multivariate} for the SAR model can be investigated.
Second, we consider the scenario in which the network data have a fixed covariate dimension, which may not be sufficiently flexible in the intricate social structure. Therefore, developing a distributed estimation method for high-dimensional data still needs to be studied.
Third, if we can collect the time series data of the responses, we could extend the proposed DNLSA method to a dynamic SAR model for large-scale networks.
Consequently, the proposed methodology can be applied to more diverse applications.

\section{ACKNOWLEDGEMENT}\label{sec:thanks}
The authors thank the two anonymous referees, the Associate Editor and the Co-Editor Serena Ng for their very helpful and constructive comments and suggestions on an early version of this paper.
Xuening Zhu's research is supported by the National Natural Science Foundation of China (nos. 72222009, 71991472, 12331009), Shanghai International Science and Technology Partnership Project (No. 21230780200), Shanghai B\&R Joint Laboratory Project (No. 22230750300), MOE Laboratory for National Development and Intelligent Governance, Fudan University, IRDR ICoE on Risk Interconnectivity and Governance on Weather/Climate Extremes Impact and Public Health, Fudan University.
Hansheng Wang's research is supported by the National Natural Science Foundation of China (nos. 12271012).
\newpage

\bibliographystyle{chicago}
\bibliography{xuening}

%
%



	
\end{document}